%% file: honeycomb_arxiv.tex
\documentclass[letterpaper,twocolumn,10pt]{article}
\usepackage{usenix-2020-09}

\usepackage{hyperref}
\usepackage[noabbrev]{cleveref}

\usepackage{color}
\usepackage{verbatim}
\usepackage{todonotes}
\usepackage{enumitem}
\usepackage{url}
\usepackage{xspace}

\usepackage{makecell}

\usepackage{caption}

\usepackage{subfig}

\usepackage[symbol]{footmisc}

\newcommand{\code}[1]{\texttt{#1}}

\newcommand{\sys}{Honeycomb\xspace}

\begin{document}

%%
%% The "title" command has an optional parameter,
%% allowing the author to define a "short title" to be used in page headers.
\title{\sys: ordered key-value store acceleration on an FPGA-based SmartNIC}

%for single author (just remove % characters)
\author{
{\rm Junyi Liu}\\
Microsoft
\and
{\rm Aleksandar Dragojevi\'{c}\footnotemark[1]}\\
Citadel Securities
\and
{\rm Shane Flemming\footnotemark[1]}\\
AMD
\and
{\rm Antonios Katsarakis\footnotemark[1]}\\
Huawei Research
\and
{\rm Dario Korolija\footnotemark[1]}\\
ETH Zurich
\and
{\rm Igor Zablotchi\footnotemark[1]}\\
Mysten Labs
\and
{\rm Ho-cheung Ng\footnotemark[1]}\\
Imperial College London
\and
{\rm Anuj Kalia}\\
Microsoft
\and
{\rm Miguel Castro}\\
Microsoft
% copy the following lines to add more authors
% \and
% {\rm Name}\\
%Name Institution
} % end author

\date{}
\maketitle

\footnotetext[1]{This work was done when affiliated with Microsoft}

\footnotetext[2]{This work has been submitted to the IEEE for possible publication. Copyright may be transferred without notice, after which this version may no longer be accessible.}

%%
%% The abstract is a short summary of the work to be presented in the
%% article.
\input{abstract.tex}

\section{Introduction} \label{sec:intro}
% Opening

%TODO: change story from pcie bottleneck to off-chip bandwidth bottleneck?
%TODO: log block without reserved space

In-memory key-value stores are an important building block in modern distributed applications~\cite{redis,memcached,tao,azurestorage,a1}. They store data in servers that provide a simple interface ({\sc put(K,V)}, {\sc V=get(K)}, {\sc update(K,V)} and {\sc delete(K)}) to clients across the network. Ordered key-value stores expand the set of supported applications
%~\cite{redis,tao,azurestorage,a1} 
by providing an efficient {\sc scan} operation to retrieve the key-value pairs whose keys are within a specified range.
For example, distributed file systems can use {\sc scan} to map ranges of logical file offsets to the nodes storing the data~\cite{azurestorage}. It is also used to query graph stores~\cite{tao,a1} and the popular Redis~\cite{redis} offers sorted sets.
 We describe \sys, a system that provides hardware acceleration for an in-memory ordered key-value store.

There is a large body of research on improving the performance of ordered key-value stores, e.g.,~\cite{masstree,erpc,cell,farmsosp,drtm,ziegler,learnedcache}. Recent research has shown how to leverage modern networks to achieve high throughput and low latency with an RPC-based system~\cite{erpc}. Other research has explored using one-sided RDMA reads to bypass the server CPU for {\sc get} and {\sc scan} operations~\cite{farmsosp,drtm,cell,ziegler,learnedcache}. Since RDMA NICs only provide simple one-sided reads of contiguous memory regions, these systems require at least two RDMA reads per operation when supporting variable-sized keys or values, and they require client-side caching to avoid additional RDMAs when traversing the tree data structures stored by the servers.

\sys accelerates an ordered key-value store using an FPGA-based SmartNIC~\cite{Putnam2014,Caulfield2016} attached to a host CPU. 
These SmartNICs are widely deployed in data centers~\cite{Putnam2014,Caulfield2016,firestone2017vfp,firestone2018azure,fowers2018a}. They enable effective CPU offload by avoiding the functionality limitations of RDMA and the performance problems of SmartNICs based on general-purpose, low-power cores~\cite{prism}. Previous work used a similar SmartNIC to accelerate an unordered key-value store~\cite{kv-direct}. \sys solves a harder problem because ordered key-value stores use more complex data structures whose operations perform O(log(n)) memory accesses instead of the O(1) accesses performed by hash table operations in unordered key-value stores.

\sys implements the ordered key-value store using a B-Tree~\cite{btree} that supports variable-sized keys and values that are stored inline in B-Tree nodes to improve the performance of scans. It uses
multiversion-concurrency control (MVCC~\cite{bernstein1987}) to provide
linearizability~\cite{linearizability} for scans. This is important to provide strong consistency, e.g., in the distributed file system example above, clients may fail to read previously written data when scans are not linearizable.

\sys stores the B-Tree in host memory. It accelerates {\sc get} and {\sc scan} operations by executing them on the FPGA, but {\sc put}, {\sc update} and {\sc delete} operations are executed by the host CPU. Since most workloads are read-dominated~\cite{memcachedanalysis,tao,bwtree,ycsb,facebookrocksdb},
there is less benefit in accelerating write operations and the cost of accelerating them in hardware is high because they may require complex splitting and merging of B-Tree nodes. This hybrid approach enables much larger stores than would be possible using the limited DRAM attached to the FPGA and it reduces the complexity of the FPGA implementation. However, it brings the challenge of data access and synchronization across the slow PCIe bus. We overcome this challenge with careful co-design of the software running on the CPU and the FPGA hardware image.

\sys uses several techniques to reduce data accesses over PCIe. It caches B-Tree nodes on both FPGA SRAM and on-board DRAM. It also uses large B-Tree nodes with {\em shortcuts}, a list of sorted keys and offsets into the node that divide it into segments of similar size. This allows requests to fetch only the relevant segments of each node they access, which reduces the number of bytes accessed to search for a key and the total amount of memory required to cache interior nodes compared to simply using smaller nodes.

To use available off-chip FPGA bandwidth efficiently, \sys exploits request-level parallelism and avoids head of line blocking with out-of-order request execution. To use all available off-chip bandwidth, \sys also implements a dynamic load balancer that directs some requests that hit on the cache to host memory if there is no bandwidth available to on-board DRAM and there is unused bandwidth over PCIe.

\sys reduces the cost of synchronization across PCIe by making {\sc get} and {\sc scan} operations wait free~\cite{waitfree}.
It also divides each leaf B-Tree node into a {\em sorted block} with sorted key-value pairs and a small {\em log block} that logs recent {\sc put} and {\sc delete} operations. The log block is merged into the sorted block when its size excedes a threshold. This ensures reads traverse mostly sorted items while avoiding 
sorting the node  and synchronizing the CPU and the FPGA on each write. 

\if 0
\sys also introduces the {\em release ring}, a batching technique to reduce communication across PCIe when making new node versions visible to readers with MVCC.
\fi

\if 0
\begin{figure}[t]
    \centering
    \includegraphics[width=0.8\columnwidth]{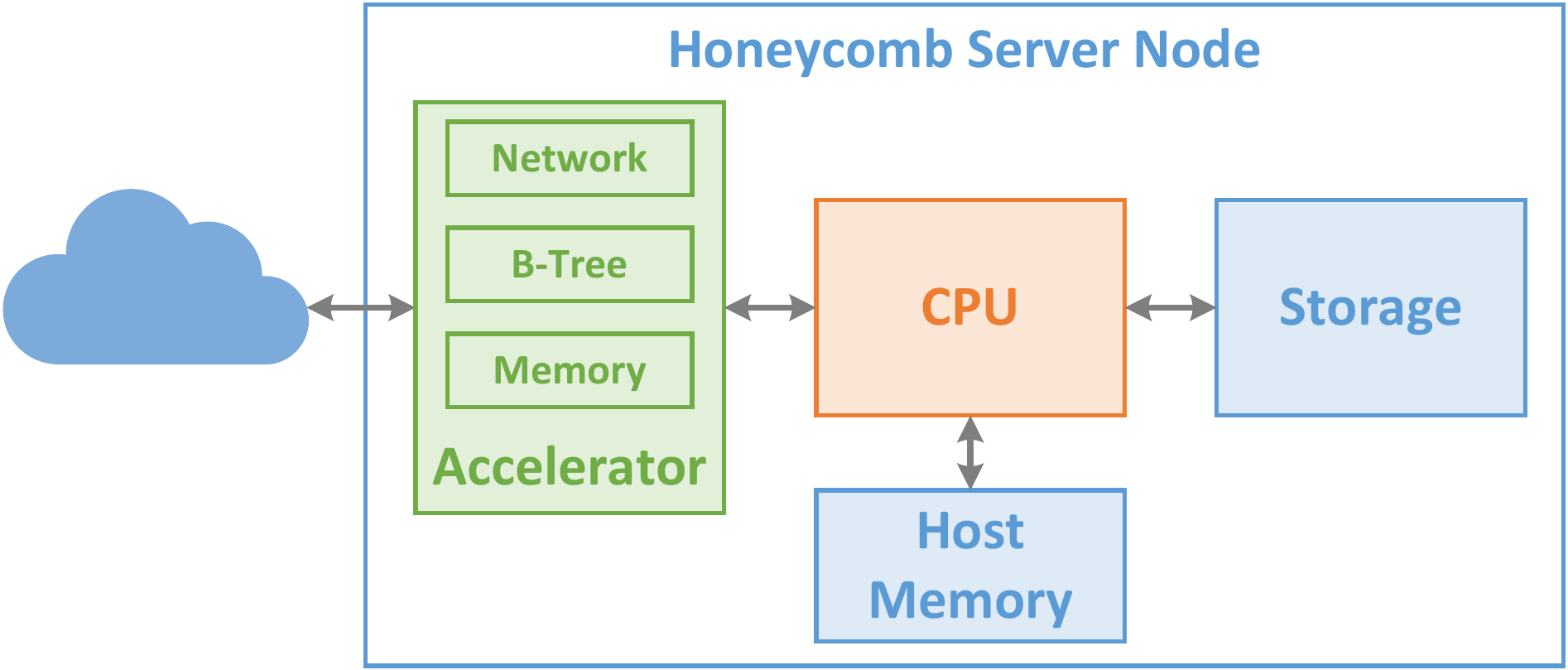}
    \caption{Block diagram of \sys B-Tree server node.}
    \label{fig:server_node}
\end{figure}
\fi

A comparison with a state-of-the-art ordered key-value store, which uses eRPC~\cite{erpc} and Masstree~\cite{masstree}, shows that \sys improves both peak performance and cost-performance, which is more important in large-scale data centers~\cite{jouppi2021ten}, significantly for scan-dominated workloads.
For example, in uniform workloads with inserts and short scans, \sys improves the throughput by $1.2\times$ for workloads with 50\% reads and by more than $2.3\times$ for workloads with at least 80\% reads. Most importantly, \sys improves throughput per watt of Termal Design Power (TDP), which is a proxy for total cost of ownership~\cite{jouppi2021ten}, by more than $1.9\times$ for workloads with at least 80\% reads.
We expect these gains to increase with future hardware because newer FPGAs have more on-chip memory to cache B-Tree nodes and they are connected to the host using PCIe Gen5 that has four times higher bandwidth than the PCIe Gen3 used in our system.

\if 0
\begin{figure}[t]
    \centering
    \includegraphics[trim={0 8.7cm 0 10cm}, clip, width=\columnwidth]{figures/perf_watt_comparison.pdf}
    \caption{\sys B-Tree perf/W comparison of three server platforms.}
    \label{fig:perf_watt_comparison}
\end{figure}
\fi

We describe the architecture of \sys in \Cref{sec:hw_impl}, the algorithms and data structures used to implement the B-Tree in \Cref{sec:btree_design}, and how these are implemented in hardware in Sections \ref{sec:hw_B-Tree} and \ref{sec:hw_mem_subsys}. \Cref{sec:eval} presents the evaluation.

\section{\sys architecture}\label{sec:hw_impl}

\begin{figure}[t]
    \centering
    \includegraphics[width=\columnwidth]{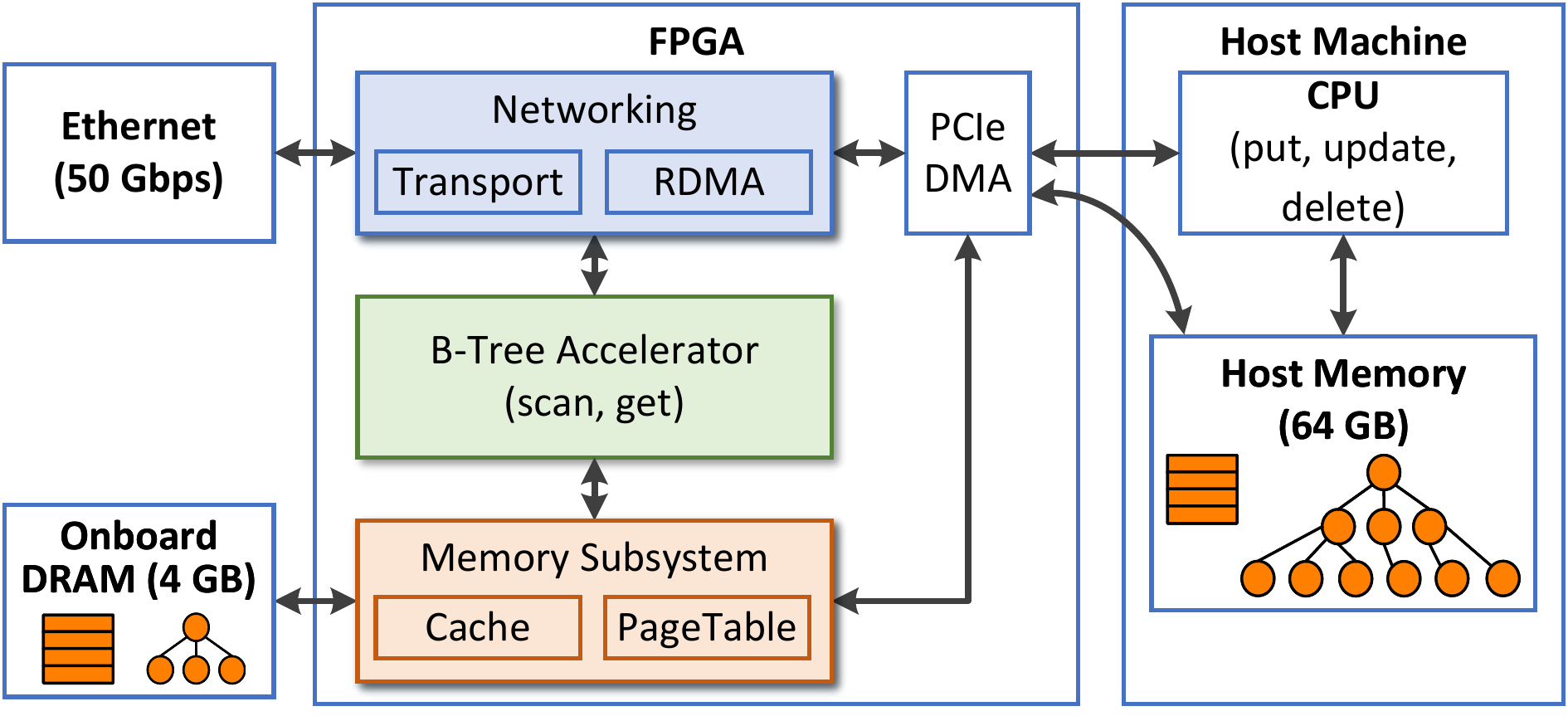}
    \caption{Hardware system architecture.}
    \label{fig:hardware_sys}
\end{figure}

\Cref{fig:hardware_sys} shows the architecture of \sys: an FPGA-based SmartNIC that connects a host server to the datacenter network.
The \sys accelerator is built on the Catapult FPGA infrastructure \cite{Putnam2014,Caulfield2016}, which provides a shell with IO support. The core FPGA image has three subsystems: the B-Tree accelerator, networking, and memory. 

The B-Tree is stored in host DRAM because it scales to much larger capacities than on-board DRAM in FPGAs. The B-Tree accelerator implements {\sc get} and {\sc scan} operations and {\sc put} and {\sc delete} operations are executed by the CPU. 

% network
The networking subsystem implements LTL~\cite{Caulfield2016}, a UDP-based transport protocol similar to RoCE v2~\cite{zhu2015congestion} on top of 50~Gbps Ethernet, and an RDMA engine to enable efficient host-to-FPGA communication. The RDMA engine supports custom commands for {\sc get} and {\sc scan} in addition to
general commands like {\sc read}, {\sc write}, {\sc send}, and {\sc receive}. These enable kernel bypass at the clients and at the servers for {\sc put}, {\sc update} and {\sc delete}. They also enable the B-Tree accelerator to completely bypass the CPU for {\sc get} and {\sc scan} requests by receiving these commands directly from the network subsystem, processing them, and sending back replies. 

% external memory
The memory subsystem interfaces with host memory over PCIe and with on-board DRAM. It maintains a cache and a page table. The cache stores interior B-Tree nodes in on-board DRAM and the root of the B-Tree in on-chip SRAM. The page table simplifies atomic updates to the B-Tree (\Cref{sec:insert}) by adding indirection~\cite{bwtree}. It is stored in on-board DRAM and maps logical identifiers (LIDs) for B-Tree nodes to their physical addresses in host memory. 

The FPGA is connected over two PCIe Gen 3 $\times8$ to the host. We measured a peak throughput of 13 GB/s and a latency of more than 1~$\mu$s (depending on load). For comparison, the bandwidth between host CPU and DRAM is up to 64 GB/s with an order of magnitude lower latency, and CPU cores have large caches that improve performance further. Therefore, \sys must address the challenge of data access and synchronization across the PCIe bus to achieve performance and cost-performance gains relative to CPU-only systems. The next sections describe how we overcame this challenge with careful hardware-software co-design.

%TODO: add page table in host to figure and say host implements put and delete

\section{B-Tree}\label{sec:btree_design}
\sys implements a B+ tree~\cite{btree}, i.e., key-value pairs (also called items) are stored only in leaf nodes while interior nodes store keys and pointers to child nodes. 
%Each leaf stores pointers to its siblings to improve the performance of scans. 
\sys guarantees linearizability~\cite{linearizability} for all operations including scans.
% This section describes the design of the \sys B-Tree. 

 \if 0
Its implementation employs fine-grained synchronization with no locking by read operations. 
The B-Tree guarantees linearizability \footnote{Each operation appears to take place atomically at a single, indivisible point in time, even if operations execute concurrently. } for all operations.
It starts from B-Tree node layout optimized for both software and hardware implementation. 
Then we discuss how the synchronization is optimized to be light-weight and the implementation details of B-Tree operations.
\fi

\subsection{Node layout} \label{sec:node_layout}
% LID and page table
B-Tree nodes are fixed-size, which is configured to 8~KB in this paper.
They are allocated in pinned host memory to allow 
the FPGA to use physical memory addresses directly.
B-Tree nodes do not store addresses of child or sibling nodes directly. 
Instead, they store 6-byte logical identifiers (LIDs)~\cite{bwtree}.  
The mapping from a LID to the virtual and physical addresses of a node is maintained in a page table. 
% With 8-byte table entries stored on 4-GB on-board DRAM, we support trees up to 4 TB, which is large enough for in-memory stores on modern servers. 

\if 0
, allowing us to store nodes on larger storge devices such as NVMe drives in the future. 
It also helps with synchronizing accesses to the tree, as write operations perform copy-on-write of nodes in some cases (more details in \textcolor{red}{\Cref{sec:insert}}). 
\fi

\begin{figure}[t]
    \centering
    \includegraphics[width=0.75\columnwidth]{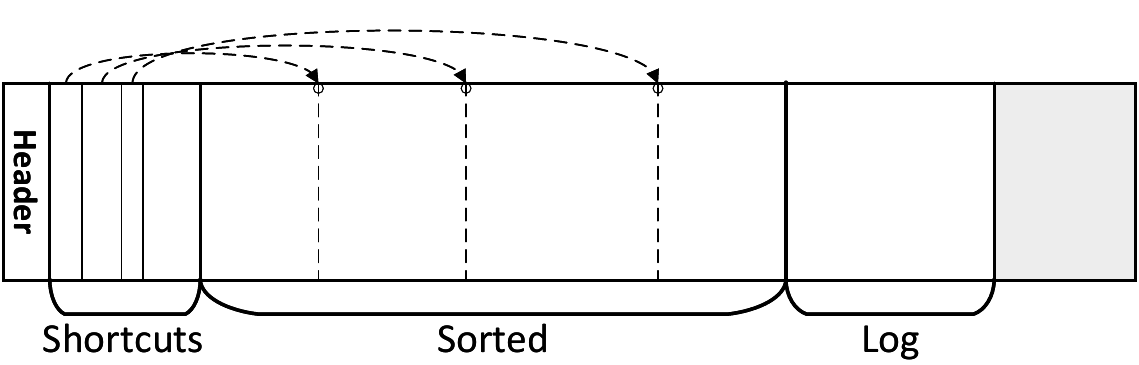}
    \caption{\sys node layout.}
    \label{fig:node_layout}
\end{figure}

% Node header and item
The node layout is shown in \Cref{fig:node_layout}. 
The node header is stored in the first 48 bytes. 
It contains fields specifying the node type (interior or leaf), the number of bytes used, a lock word, and a version number to synchronize accesses to the node. The header of an interior node stores the LID of the leftmost child. 
The header of a leaf node stores the LIDs of the siblings to improve scan performance. Leaf nodes store keys and values in variable-size blobs. 
Each blob has a 2-byte header that specifies its size. 
Keys and values are stored inline in B-Tree nodes to improve performance.
In the current configuration, the maximum key length is 460~bytes and values larger than 469~bytes are stored outside the node.

% Intro large and small block
Leaf nodes are split in two blocks: sorted and log. 
The pointer to the boundary between them is stored in the node’s header. 
%The pointer is an offset relative to the start of the large block. 
The sorted block stores items in ascending lexicographic order of keys and the log block stores a log of recent changes. 
When the log block grows larger than a threshold (currently set to 512 bytes), it gets merged with the sorted block. 
Therefore, read operations benefit from accessing mostly sorted data and write operations avoid the overhead of sorting the node on every change. It also reduces synchronization across PCIe because it requires one update to the FPGA page table per-merge instead of one on every write operation. 

Since interior nodes change infrequently, the cost of storing a log block and accessing it on reads outweighs the benefits. Therefore, interior nodes do not have log blocks.

% Small block
Entries in the log block are either newly inserted or updated items or delete markers. 
Each entry stores a 2-byte pointer to an item in the sorted block. 
For a newly inserted item, it points to the first item with key greater than the item’s key. For other entries, it points to the deleted or updated item.
In addition, each entry records the 1-byte index in the sorted order of log items at the time of insertion. 
These fields are used to sort all node items, when searching the leaf, without comparing keys and with O(1) cost per log block item (\Cref{sec:small_block_sort}).

\if 0
These indexes are “replayed” as shifting and storing elements in a small indirection array, while searching the small block to rebuild the current sorted order (\textcolor{red}{more details in \Cref{sec:small_block}}). 
This sorting is resource-efficient in hardware and does not introduce significant latency. 
\fi

% Shortcut block

To optimize the search for a key in a node, we split the sorted block into segments of roughly equal size. 
We store the keys at the boundaries of segments and the pointers to the boundaries in the shortcut block, which is stored immediately after the node header. We currently reserve 48~bytes for the header and 464~bytes for the shortcut block. The shortcut keys are selected during the merge of the sorted and log blocks. They are stored only in the shortcut block, i.e., the start of the segment for boundary key K contains the value associated with K rather than another copy of K followed by the value.

The search for a key starts by traversing the shortcut block to identify the segment that contains the key. 
The search examines only that segment and the log block, which reduces the amount of data read both from host memory and the cache.
Since \sys's bottleneck is FPGA off-chip bandwidth, this optimization significantly improves performance. 
For small keys with the current configuration, a search reads at most 1.5~KB of data from an 8~KB node, which is a $5\times$ performance improvement relative to reading the whole node.

Analysis shows that using large nodes with shortcuts reduces the number of bytes accessed to search for a key and the total amount of memory required to cache interior nodes compared to simply using smaller nodes. For the example of a five-level tree with 8~KB nodes, 16-byte keys and values at 55\% occupancy, searching the tree with this optimization accesses fewer than 75\% of the bytes accessed searching a simple tree with 512-byte nodes with a similar number of items. It also requires approximately four times less space to cache all the interior nodes.
The intuition is that the FPGA can use the information in the header to fetch a sorted block segment or a log block with a single DMA that reads only the bytes in use, whereas it would need to read the whole node or issue more dependent DMAs for the simple tree. Also using smaller nodes increases overhead because it requires more storage for pointers to children in interior nodes. Using log blocks reduces the impact of sorting the large node on writes.

\subsection{Synchronization}\label{sec:sync}

{\sc get} and {\sc scan} operations executed by the accelerator are wait-free~\cite{waitfree} --
they never block or retry due to {\sc put} or {\sc delete} operations executed by the CPU.
This reduces the overhead of synchronization over PCIe and provides predictable latency.

% MVCC with two global versions
We use multi-version concurrency control (MVCC)~\cite{bernstein1987}
to ensure that read operations can always access a consistent snapshot of data. 
\sys maintains two shared 64-bit version numbers: the {\em global write version} and the {\em global read version}.
A write operation performs an atomic fetch and add on the global write version to obtain its write version, which is used to version any items it creates.
Writers {\em release changes} to readers in version order: a writer sets the global read version to its write version when it becomes the writer with the smallest write version, and updates the copy of the global read version maintained by the accelerator over PCIe. We delay responses to write requests until this update completes.
Writers synchronize with each other using node locks, which are ignored by readers. 
Read operations read the global read version maintained by the accelerator to get their read version.
They ignore items with versions greater than this value.

\if 0
By using two global version numbers, read operations always access items for which the commit is complete, without having to wait for concurrent operations. 
This means that reads are wait-free, and in our system, the hardware readers will not need to wait for software writers. 
We opt to use two global versions at the cost of ordering the release of write operations. 
This overhead is reduced by using a release ring data structure that allows efficiently determining the release order \textcolor{red}{(more details in \Cref{sec:rel_ring})}. 
\fi

% Node and item versioning
All items in the sorted block have the same version that is stored in the node header and is called the \textit{node version}. 
The node version is the write version of the write operation that created the version of the node. Nodes also store a pointer to the state of the node as of the previous version. A reader that observes a node version greater than its read version follows the chain of old version pointers until it reaches a node with a version less than or equal to its read version. The version of an item in the log block is stored in the item header.
To reduce the size of these headers, we store a 5-byte version delta from the node version instead the entire 8-byte version. 
If a write would cause the delta version to wrap around, it merges the sorted and log blocks. 
Readers ignore all items in the log block with versions greater than their read version. 

% Node garbage collection, garbage collection of old versions
To ensure operations observe a consistent snapshot, \sys does not reclaim old versions of nodes while at least one inflight operation can access a snapshot to which the old version belongs. To garbage collect old versions that are not accessible anymore, \sys uses an epoch-based memory manager~\cite{mckenney2007rcu}. Each CPU thread executes operations one at a time, assigns a sequence number to each inflight operation, and exposes its current operation sequence number to the memory manager. Similarly, the B-Tree accelerator assigns a sequence number to each operation and exposes the sequence numbers of the newest and the oldest inflight operation to the memory manager over PCIe. When a CPU thread makes a change that removes old versions of nodes from the tree, it puts the old versions into a garbage collection list tagged with a vector timestamp. This vector timestamp has entries for the current operation sequence numbers of all CPU threads and the newest operation sequence number on the accelerator. 

Periodically, a thread scans the garbage collection list and reclaims all node versions that are not reachable by CPU threads or the accelerator, i.e., node versions with a vector timestamp where the entry for each CPU thread is smaller than the current operation sequence number for the thread, and the entry for the accelerator is smaller than the current oldest inflight operation on the accelerator. If a writer fails to allocate memory for a new node and the garbage collection list is not empty, the operation is aborted and retried.
 
The page table is stored both in host DRAM and in FPGA on-board DRAM. 
% (but the copy in the FPGA only stores physical addresses).
When a new node mapping is created or when a node’s mapping is changed, the CPU updates the table in host memory and issues a PCIe command to the FPGA to update the copy of the table in FPGA on-board memory. 

\sys can be configured with MVCC off by using zero as the version number for all nodes and log items, and by not setting old version pointers. This improves write performance a little by avoiding updates to the global read version on the FPGA.
The atomic tree updates using  the page table still ensure linearizability for {\sc get}, {\sc put}, and {\sc delete}, but not for {\sc scan} because sibling pointers are not updated atomically. 

\if 0
 Many ordered key-value stores offer only this weaker guarantee (e.g.,~\cite{masstree}), which is sufficient for many applications. 
\fi

\subsection{Scan and get} \label{sec:lookup} \label{sec:scan}

\sys implements {\sc scan(K$_l$, K$_u$)} that finds the largest key {\sc K$_s$} less than or equal to {\sc K$_l$}, and returns a sorted list with all key value pairs with keys greater than or equal to {\sc K$_s$} and less than or equal to {\sc K$_u$}. We implement this version of {\sc scan} to support mapping from a range of logical file offsets to the nodes storing the data in distributed file systems~\cite{azurestorage}. In this application, the key-value pairs represent variable-sized chunks of file data where the key is the logical offset of the first byte in the chunk and the value points to the servers storing the data. Since the start of a logical file offset range in a client request can be in the middle of a chunk, the  {\sc scan} semantics we implement are necessary for correctness.

{\sc scan(K$_l$, K$_u$)} traverses the tree from the root to a leaf. It fetches the node header and shortcut block of each node it visits (currently stored in the first 512 bytes). It follows old version pointers if needed to find a node with version less than or equal to the operation's read version. Then it searches for the largest shortcut key less than or equal to {\sc $K_l$}, and fetches the associated sorted block segment. When {\sc scan} visits an interior node, it searches the segment for the item with the largest key less than or equal to {\sc $K_l$} (which could be the shortcut key), and obtains the LID of the next node to visit from the item.

When {\sc scan} reaches the leaf, it  fetches the log block in parallel with searching the shortcuts. Then it searches both the sorted and log blocks for the largest key {\sc K$_s$} less than or equal to {\sc K$_l$}. It scans forward from {\sc K$_s$} until it reaches a key greater than {\sc K$_u$} or the end of the tree. It ignores any items with versions greater than the read version. If it finds one or more items with the same key, it returns the value associated with the item with the latest version unless that item is a delete. Otherwise, it ignores the item. {\sc scan} uses sibling pointers to move to the next leaf if needed.

The accelerator optimizes scan performance by determining a sorted order for the items in the log block of a leaf, and uses the back pointers from log block items to sorted block items to determine a sorted order across all the items in the leaf (see~\Cref{sec:small_block_sort}). This enables efficient scanning by overlapping data access with key comparisons, generating results that are already sorted, and avoiding key comparisons with all items in the log block.

\sys implements {\sc get(K)} as {\sc scan(K,K)} and post-processes the result to return the value associated with {\sc K}, or return not found if  {\sc K} is not in the result.

\if 0
XXX don't think we need this either.
To keep the items sorted, the scan of a leaf merges the items in the range from the three node blocks.
 
While scanning, the ordering between the large and the small block is preserved by the back pointer in the small-block item. 
If the next small-block item points to the next large-block item, the small-block item is returned and the scan moves to the next small-block item; otherwise, the large-block item is returned, and the scan moves on to the next large-block item. 
To handle shortcut keys correctly, the scan keeps track of the end of the current large-block segment. 
At the beginning of each segment, the scan retrieves the key from the shortcut block and the value from the large block. 
% In the middle of a segment, both the key and the value are in place. 
\fi
\if 0
XXXX from the old lookup; I don't think we need this.
Items in the small block with a version newer than the lookup’s read version are ignored. 
The lookup follows the pointer stored in the item with the larger key that found in the small and the large block. 
Making this decision does not require comparing the keys – if the back pointer of the small block item points just after the large block item, it follows the pointer in the small block; otherwise, it follows the pointer in the large block. 
If the target key is smaller than both the first key in the large block and in the small block, the lookup follows the leftmost pointer. 
XXXX
\fi

\subsection{Insert} \label{sec:insert}

A {\sc put(K,V)} operation inserts a new item if there is no item with key {\sc K} in the tree. It traverses the tree as described for {\sc scan(K,K)} without acquiring any locks, but it reads the latest version of each node to implement linearizability. The operation locks the leaf node before modifying it to ensure the state matches the one observed during the traversal.

The node header has a 32-bit lock word with a lock bit and a 31-bit sequence number, which is incremented on writes to the node.
{\sc put} tries to acquire the leaf's lock using an atomic compare-and-swap instruction that checks if the sequence number matches the one observed during the traversal. If the compare-and-swap fails, the {\sc put} operation restarts. 

{\bf Fast-path insert:}
In the common case, the operation simply adds a new item to the log block: it copies the item to the log block; acquires the write version with a fetch-and-add on the global write version; sets the version of the item in the log block; updates the node size; increments the node sequence number; and unlocks the node. The node size and the lock word are packed into a 64-bit word. So that the last three steps can be performed with a single instruction. Concurrent operations either observe the node without the item or with the item correctly written, and readers ignore the item until the change is released (\Cref{sec:sync}).

%They also ignore the item before its write version is released, as the read version is older than the item. 
% Note that FPGA does not cache leaf nodes. If it did, the writer would have to invalidate the cache by issuing commands over PCIe, which would introduce additional overheads to the common case insert.

\begin{figure}[t]
    \centering
    \includegraphics[width=\columnwidth]{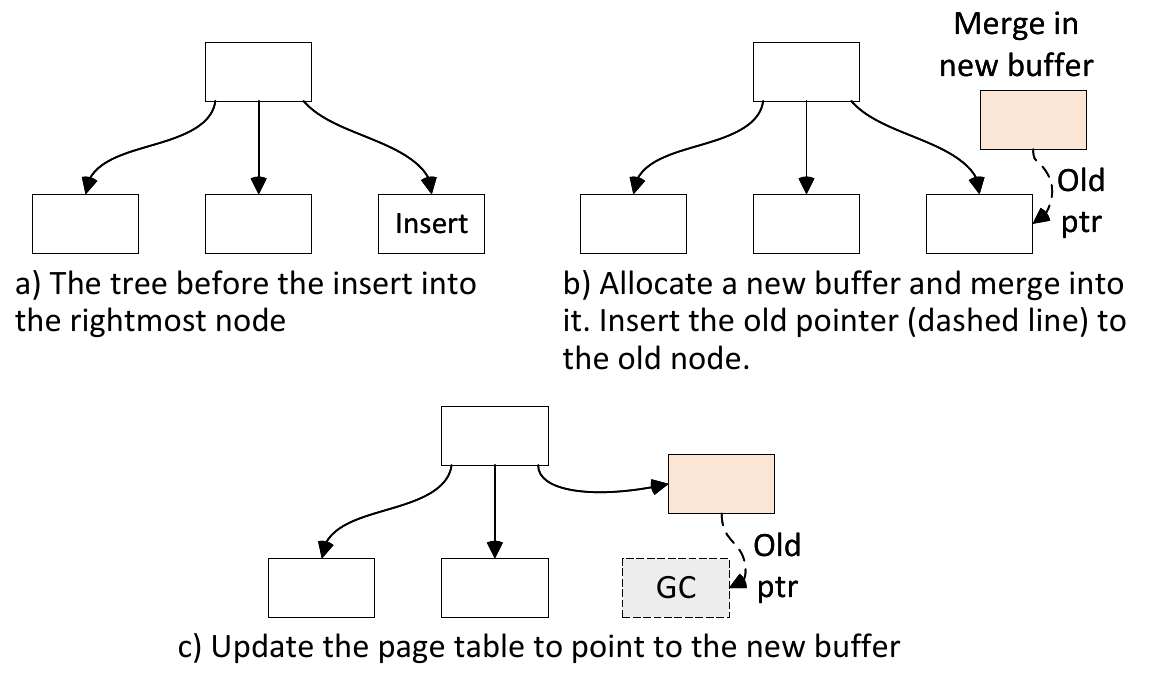}
    \caption{Merging the sorted and log blocks.}
    \label{fig:large_small_merge}
\end{figure}

{\bf Sorted and log block merge:}
If the size of the log block exceeds a threshold (currently 512 bytes),
the operation merges the sorted and log blocks as shown in \Cref{fig:large_small_merge}. 
It allocates a new memory buffer for the node and sorts all the items into a single sorted block in the new buffer.
While sorting, it selects the shortcut keys.
For each item, it decides whether to put the key in the shortcut block based on the number of bytes copied so far, the remaining bytes to copy, and the average size of keys and values. It attempts 
to maximize the number of shortcut keys while ensuring that segments have similar sizes, a minimum size (currently 256 bytes), and that 
shortcut keys fit in the shortcut block.

The operation sets the node version to the write version, and the old version pointer to the address of the old buffer (\Cref{fig:large_small_merge}b). Then it updates the LID mapping with the address of the new buffer in the page tables of both CPU and FPGA (\Cref{fig:large_small_merge}c). The parent node does not need to be changed because the LID of the child remains the same. The operation also puts the old buffer in the garbage collection list and releases the changes to make them visible to readers.
This ensures concurrent operations observe the change atomically.

\if
, unlocks it, and sets the “deleted” flag in its header to ensure that concurrent writers do not update the old buffer. 
Ongoing operations can still read from the old buffer, but if an operation needs to update a deleted node, it retries. 
On retry, it will use a new read version and the new node mapping.
\fi

\begin{figure}[t]
    \centering
    \includegraphics[width=\columnwidth]{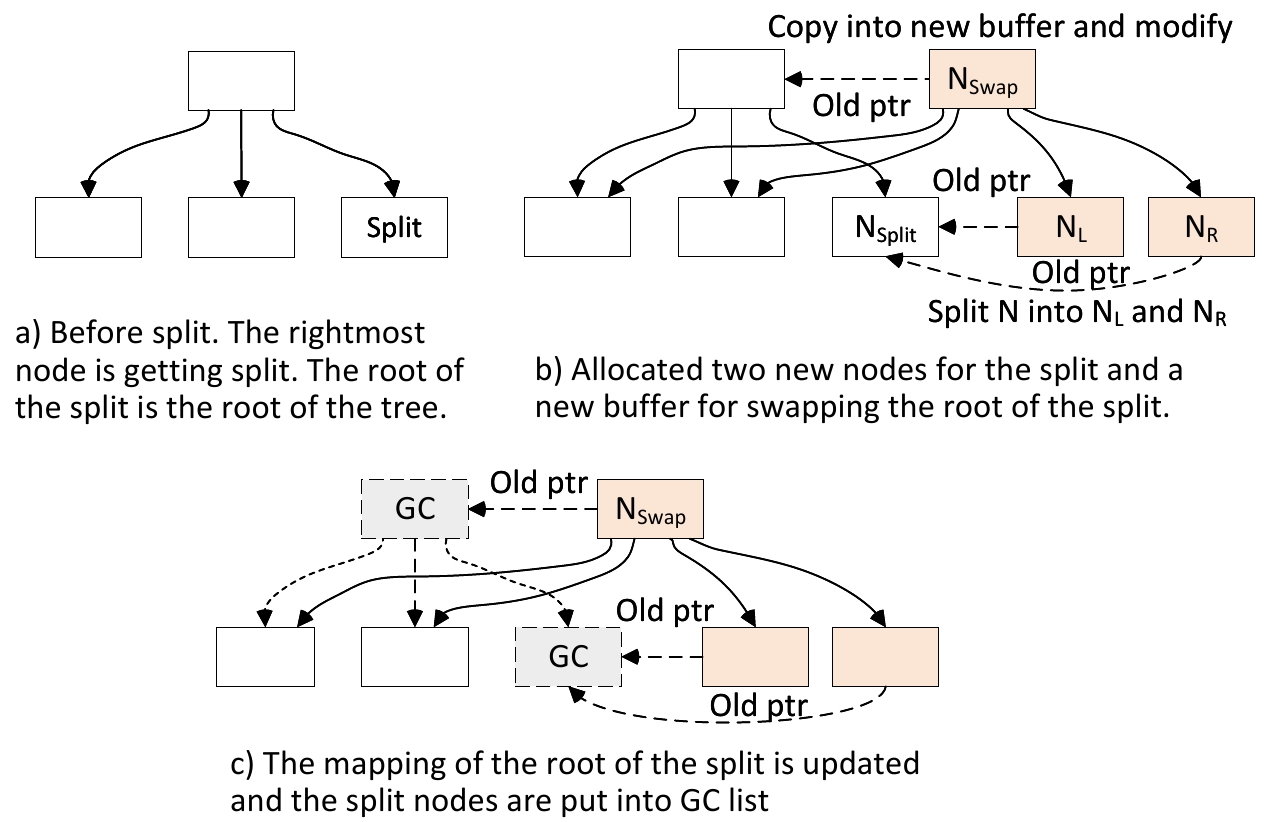}
    \caption{Splitting a node during insert. 
    %$N_{swap}$ is a newly allocated node buffer. $N_{L}$ and $N_{R}$ are new nodes with a new LID and a memory buffer.
    }
    \label{fig:node_split}
\end{figure}

% \begin{figure*}[t]
%     \centering
%     \includegraphics[width=0.9\textwidth]{figures/btree_hw_design}
%     \caption{B-Tree accelerator architecture.}
%     \label{fig:btree_hw_design}
% \end{figure*}

% \begin{figure}[t]
%     \centering
%     \includegraphics[width=0.6\columnwidth]{figures/ksu_design}
%     \caption{Key seach unit}
%     \label{fig:ksu_design}
% \end{figure}

\begin{figure*}[t]
    \centering
    \begin{minipage}[b]{0.79\textwidth}
        \centering
        \includegraphics[width=\textwidth]{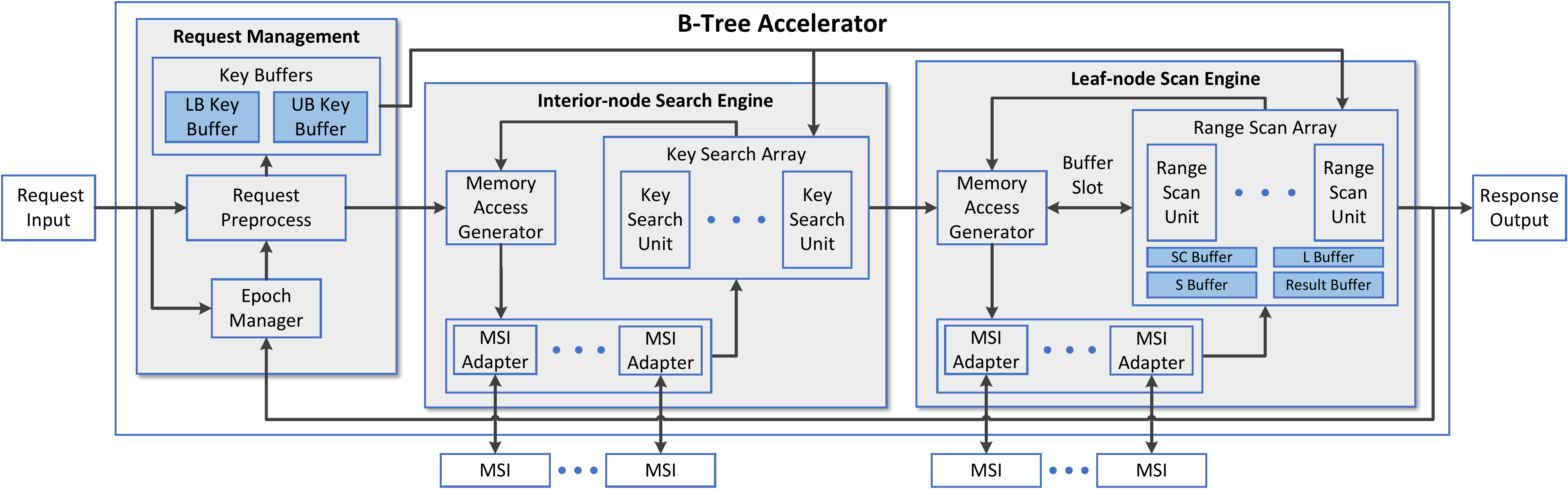}
        \caption{B-Tree accelerator architecture.}
        \label{fig:btree_hw_design}
    \end{minipage}
    % \hfill
    % \hspace{3pt}
    \begin{minipage}[b]{0.195\textwidth}
        \centering
        \includegraphics[width=\textwidth]{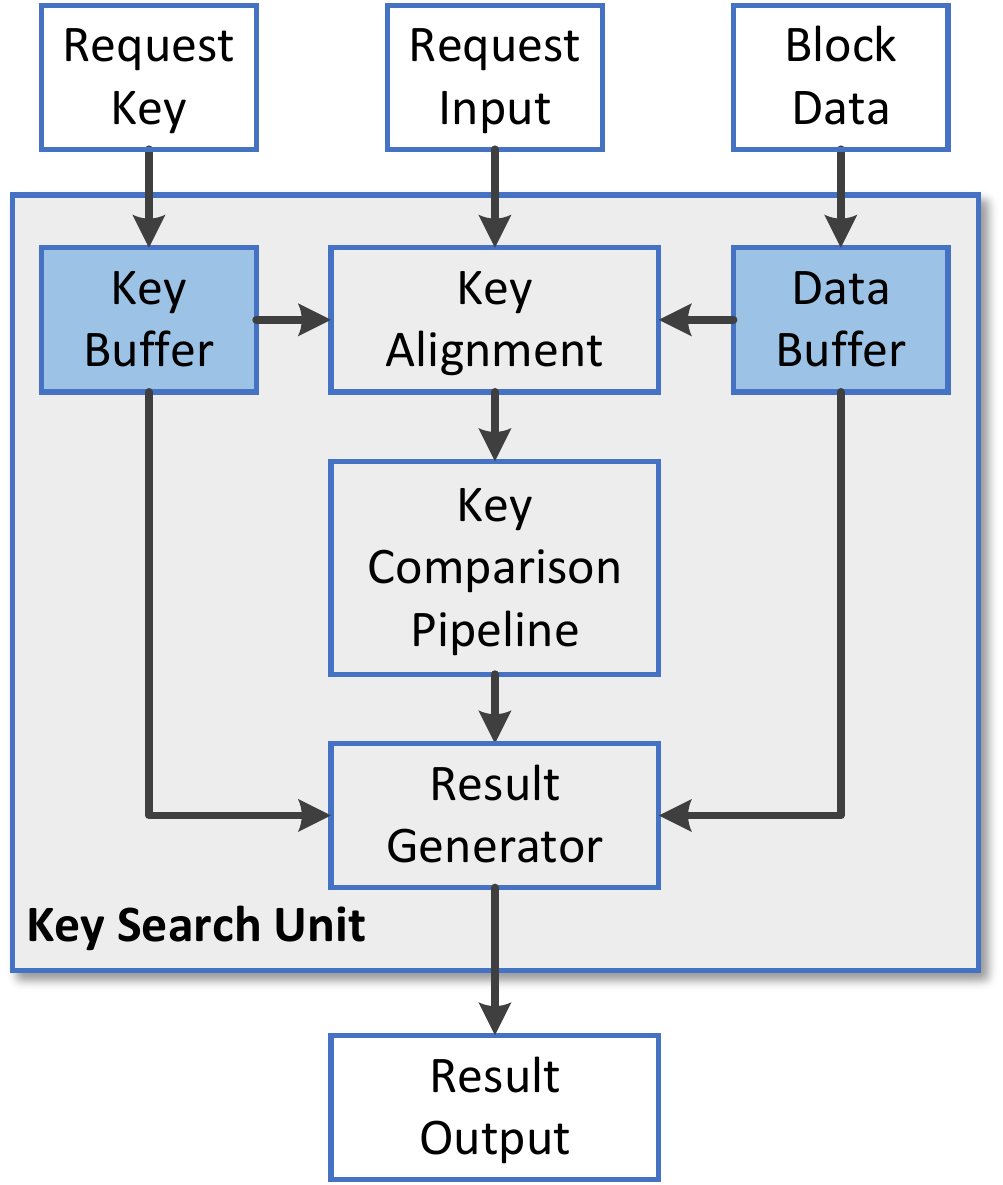}
        \caption{Key seach unit}
        \label{fig:ksu_design}
    \end{minipage}
    \vspace{-6pt}
\end{figure*}

{\bf Node splitting:}
If there is not enough space in the leaf for the merged block, the operation splits the leaf in two. This requires inserting a new item in the parent node as shown in \Cref{fig:node_split}. If the parent is full, the operation also splits the parent. Splitting can propagate to the root. The last interior node that is updated but not split is called \emph{the root of the split}. 
%The operation creates a new subtree with all the splits and replaces the old subtree atomically.

The operation acquires  locks on all interior nodes to split and on the root of the split. If locking fails due to a sequence number mismatch, it 
releases all acquired locks and restarts. The operation also allocates two new nodes (new LID and memory buffer) for each node that it splits and a memory buffer (same LID) for the root of the split ($N_L$, $N_R$, and $N_{swap}$ in \Cref{fig:node_split}b). Then it processes nodes from the leaf up splitting the items in each node evenly into the two newly allocated nodes. When it gets to the root of the split, the operation copies the items into the new buffer and adds the LIDs of the new child nodes and the key at the boundary between them. 
%If memory allocation fails because system is out of memory, the writer gives up and returns the appropriate status code to the caller.

To swap in the new subtree atomically, the operation updates the LID mapping for the root of the split in both CPU and FPGA page tables to point to the new buffer (\Cref{fig:node_split}c). It also locks the sibling leaves and updates their sibling pointers to point to the new leaves. This is sufficient to ensure linearizability for all operations but scans, because the updates to the sibling leaves and the subtree swap are not atomic. 

Linearizable scans are important for applications like the distributed file system~\cite{azurestorage}. Without them, a client $C$ could write data $D$ to a file at a logical offset range $R$, and a later read from a range containing $R$ could fail to return $D$. This can happen if the B-Tree node with the mapping for the servers that store $D$ is split and the sibling pointers are not updated before the scan to find the servers storing the data being read.

To provide linearizable scans, the operation sets the versions in all newly allocated buffers to its write version, and sets the old version pointer in the new root of the split to the address of the old buffer (\Cref{fig:node_split}b). This ensures that operations with older read versions traverse the old subtree. It also sets the old version pointers of the new leaves to the address of the old leaf. This ensures that scans with older read versions, which may reach the new leaves using sibling pointers, traverse the old leaf. The operation also puts the LIDs and memory buffers of all nodes that got split, and the old memory buffer of the root of the split in the garbage collection list. The operation then unlocks all nodes and releases the changes to readers (\Cref{sec:sync}). This ensures readers observe all changes made during the split, or none of them. 
 
If the tree root must be split, the operation grows the tree by allocating a LID and a memory buffer for the new root, and updating the root and tree height in the accelerator.
%, which is equivalent to swapping in a new subtree. 

\subsection{Delete and update}

\if 0
The implementations of {\sc delete} and insert are similar. {\sc delete} appends a delete entry to the log block, which points to the item being deleted in the sorted block to indicate that readers should ignore the item. The space for deleted items is reclaimed when the sorted and log blocks are merged. Nodes whose occupancy drops below a threshold are merged and nodes that become empty are deleted. We use similar techniques to ensure atomicity as for splits but omit the details.

A {\sc put(K,V)} operation updates the item with key {\sc K} if it is already in the tree. We implement update as a combination of delete and insert. It appends a delete entry for the old item and an insert entry for the new item to the log block atomically. 
\fi

The implementations of {\sc update} and {\sc delete} are similar to {\sc insert}.
{\sc update} appends the updated key-value pair to the log block, which points to the previous version of the item in the sorted or log block. For {\sc delete}, a delete marker is appended in the same way to indicate that readers should ignore the deleted item. The space for stale and deleted items is reclaimed when merging the sorted and log blocks. Nodes whose occupancy drops below a threshold are merged and nodes that become empty are deleted. We use similar techniques to ensure atomicity as for splits but omit the details.

\section{B-Tree accelerator implementation}\label{sec:hw_B-Tree}

% accelerator overview
The B-Tree accelerator has three components (\Cref{fig:btree_hw_design}): request management, interior-node search engine, and leaf-node scan engine. Request management parses requests and assigns them sequence numbers and read versions. The interior-node search engine is responsible for traversing the tree from the root to a leaf.  The leaf-node scan engine traverses the leaf or, for {\sc scan} operations, leaves using sibling pointers.

\subsection{Request management}
% out-of-order request execution
The time to complete requests is variable, e.g., due to different size keys or cache misses. Therefore, the B-Tree accelerator
exploits request-level parallelism and avoids head of line blocking by supporting out-of-order request execution. This 
maximizes the utilization of compute resources, and on-board DRAM and PCIe bandwidth.

% Key buffers
The request pre-process module extracts the lower (LB) and upper bound (UB) keys from the request and stores them in centralized key buffers. These buffers support multiple read ports to enable the interior-node search engine and the leaf-node scan engine to read multiple keys in parallel. When a request completes, its key buffers are freed.

The request pre-process module also initializes {\em request metadata} that includes: identifiers of the buffers holding the keys, the LID of the node being traversed, the node's level, the block type being traversed (shortcut, sorted, or log), and the start offset of the block/segment within the node. The request metadata flows over a wide data path through the accelerator.

Request management is also responsible for storing the accelerator's copy of the 64-bit global read version, which is updated 
by the CPU over PCIe (\Cref{sec:sync}). It reads the global read version to assign a read version to each incoming request. Requests are also assigned a 64-bit sequence number by incrementing a counter $S_{new}$ maintained by the epoch manager.
These are part of the request metadata.

The epoch manager also monitors request completion to keep track of $S_{old}$, the sequence number of the oldest inflight request. It exposes $S_{old}$ and $S_{new}$ over PCIe to the memory manager running on the CPU to enable reclaiming memory when it is no longer accessible by inflight requests (\Cref{sec:sync}). Accesses to $S_{old}$ and $S_{new}$ over PCIe are infrequent.

% Epoch control
%If $(S_{new} - S_{old})$ is greater than the epoch length, no new requests will be accepted by the accelerator. 

% % sync with software
% When the software makes any changes on the tree, it can record $S_{new}$ from the hardware to be $S_{change}$. 
% Once the software sees $S_{old} \geq S_{change}$, this means that all the old requests potentially accessing the staled tree nodes have been finished. 
% In other words, B-Tree accelerator can only reach the tree nodes with the new changes. 
% From this point of time, the staled tree nodes will be freed safely to recycle their memory space.

\begin{figure*}[t]
    \centering
    \begin{minipage}[b]{0.48\textwidth}
        \centering
        \includegraphics[width=\columnwidth]{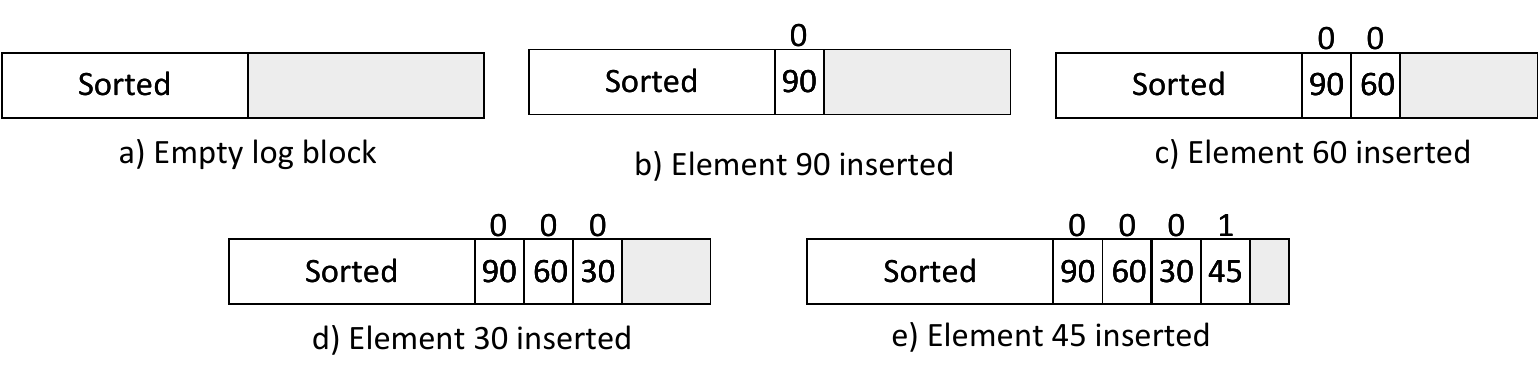}
        \caption{Example of order hints in the log block.}
        \label{fig:small_block_insertion}
    \end{minipage}
    % \hfill
    % \hspace{3pt}
    \begin{minipage}[b]{0.48\textwidth}
        \centering
        \includegraphics[width=\columnwidth]{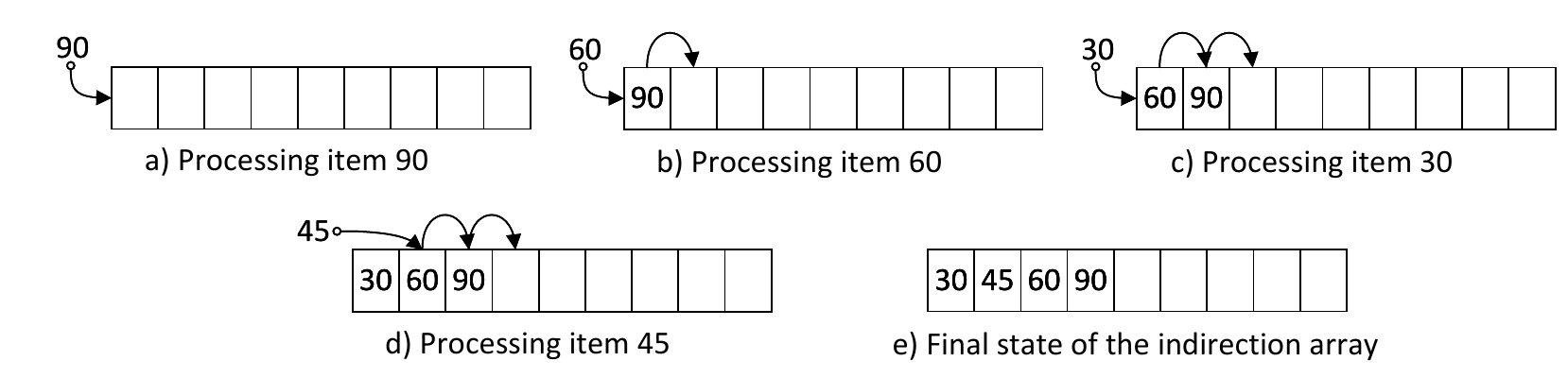}
        \caption{Sorting the log indirection array with order hints.}
        \label{fig:small_block_sort}
    \end{minipage}
    \vspace{-6pt}
\end{figure*}

% \begin{figure}[t]
%     \centering
%     \includegraphics[width=\columnwidth]{figures/small_block_insertion.pdf}
%     \caption{Example of order hints in the log block.}
%     \label{fig:small_block_insertion}
% \end{figure}

% \begin{figure}[t]
%     \centering
%     \includegraphics[width=\columnwidth]{figures/small_block_sort.pdf}
%     \caption{Sorting the log indirection array with order hints.}
%     \label{fig:small_block_sort}
% \end{figure}

\subsection{Interior-node search engine}

% overview
The interior-node search engine is responsible for traversing interior nodes from the root as described in \Cref{sec:lookup}. It uses a ring architecture to search node blocks iteratively while overlapping memory reads with compute (\Cref{fig:btree_hw_design}). The architecture exploits request-level parallelism by using multiple memory subsystem interface (MSI) adapters  and key search units (KSU). Additionally, each MSI adapter supports out-of-order data transfer to fully utilize off-chip bandwidth.

For each interior node visited by a {\sc get} or {\sc scan} request, the memory access generator first generates a memory access request to one of the MSI adapters to fetch the node's header and shortcut block. The adapter issues the request to the memory subsystem, waits for the response, and sends the response to one of the KSU. 
The KSU first checks if the version in the node header is greater than the request's read version. If so, it sends the updated request metadata back to the memory access generator to visit the old version of the node. If not, the KSU searches the shortcuts for the sorted block segment to fetch. 
Then sends the updated metadata back to the  memory access generator that generates a memory access request to one of the MSI adapters to fetch the sorted block segment. When the adapter gets the segment, it is sent to one of the KSU. The KSU searches the segment for the LID of the next node to visit. If the next node is a leaf, the request is passed on to the leaf-node scan engine. Otherwise, the request is passed to the memory access generator to continue with the child.

% search request generator and scan request generator are confusing names because they do not geneate k-v requests. They generate %requests to fetch blocks.

% KSU
The architecture of a KSU is shown in \Cref{fig:ksu_design}. It uses the request input metadata to fetch the request's lower bound key.
Then it processes the block data (header and shortcut, or sorted block segment) to find the largest key that is smaller than or equal to the request key. Since keys are variable size, KSU have to process keys sequentially. They process keys in a streaming fashion overlapping data transfer with key comparisons. The key comparison pipeline implements comparison logic equivalent to the C++ \code{memcmp()} function. It compares key fragments stored in registers with configurable width (currently 16~bytes). Key alignment uses barrel shifters to stream keys into the comparison pipeline. The result generator outputs updated request metadata.

\subsection{Leaf-node scan engine} \label{sec:rsu}
% overview
The leaf-node scan engine is responsible for traversing leaf nodes as described in \Cref{sec:lookup}. As shown in \Cref{fig:btree_hw_design}, it also uses a ring architecture to scan leaf node blocks iteratively, and exploits request-level parallelism by using multiple MSI adapters and range scan units (RSU). 

Unlike the key search array, the range scan array includes several buffers: {\em result buffer,  SC buffer, S buffer}, and {\em L buffer}. 
The result buffer is used to accumulate partial results because the leaf-node scan engine produces a sorted list of key-value pairs spread across different blocks in one or more leaves. The SC buffer, S buffer, and L buffer are used to buffer data from shortcut, sorted, and log blocks in a leaf, respectively. This enables iteration over keys in the three blocks in a sorted order to generate sorted results. The range scan array also includes different RSU variants to scan each type of block and finite state machines (FSM) that coordinate scans across the three RSU variants. The buffers are separate from RSUs to enable overlapping of compute with data buffering. The buffers are divided into request slots that also include the FSM state.
The memory access generator maintains a list of available slots. It assigns slots to requests or pushes back if no slot available.

\if 0
%TODO: Move this to the configuration part in evaluation
 
The number of buffer slots, $N_{slot}$, is equal to $T_{req} \times L_{range}$, where $L_{range}$ represents the latency of finishing one range scan on leaf nodes. 
In our implementation, $N_{slot}$ can be less than 64, and the total on-chip memory used for scan buffers is below 256 KB. 

The number of each RSU variants, $N_{RSU}$, is equal to $T_{req} \times L_{scan}$, where $L_{scan}$ is the latency of scanning one block data. 
In particular, the RSU designed for large block scanning consumes more cycles and logic resources to merge all key-value pairs detected in the corresponding leaf node. 
\fi

% RSU
The architecture of the RSU is similar to the KSU (\Cref{fig:ksu_design}) but it has two parallel key comparison pipelines to compare keys with both the lower and upper bound keys in the {\sc scan} operation. Additionally, the RSU variant responsible for log block processing must first sort the log block to enable iterating over the key-value pairs in order.

\label{sec:small_block_sort}

We leverage hardware-software co-design to optimize log block sorting.
Sorting does not compare keys and costs only $O(1)$ cycles per item. 
It uses 1-byte order hints stored in log block items by inserts. 
\Cref{fig:small_block_insertion} illustrates an example sequence of insertions into the log block. 
For clarity, the number in the item represents its key. 
The number above the item represents its order hint. 
The log block is initially empty. Item 90 is inserted first and is assigned order hint 0. 
Next, item 60 is inserted. 
Since it is the smallest item in the log block, it is assigned order hint 0. 
The order hints in existing items are not changed. 
Then, item 30 is inserted with order hint 0 since it is the smallest in the log block again. 
The final item is 45 and its order hint is 1. 

% The indexes are “replayed” while searching the small block to rebuild the current sorted order. 
% The established order is stored in an small indirection array with low memory overhead in both software and hardware.

Sorting creates an indirection array with offsets of items in the log block in ascending key order.
\Cref{fig:small_block_sort} illustrates sorting of the log block in \Cref{fig:small_block_insertion}. It shows keys instead of item offsets for clarity.
Items are processed in the order in which they are stored. When the item with order hint $i$ is processed, its offset in the log block is inserted into position $i$ in the indirection array and all the elements at positions $j \geq i$ are shifted to the right.
Since the indirection array is stored in a large shift register, insertion costs one cycle and can be overlapped with fetching the next item.
Any items with version greater than the operation's read version are filtered out.
After sorting, the RSU uses the indirection array to stream keys to the key comparison pipelines in order.

\if 0
If the upper-bound key is not found, range scan array will start a new scan on the sibling leaf node, and the current partial results will be forwarded to the result buffer. 
\fi

\section{Memory subsystem}\label{sec:hw_mem_subsys}

\begin{figure}[t]
    \centering
    \includegraphics[width=\columnwidth]{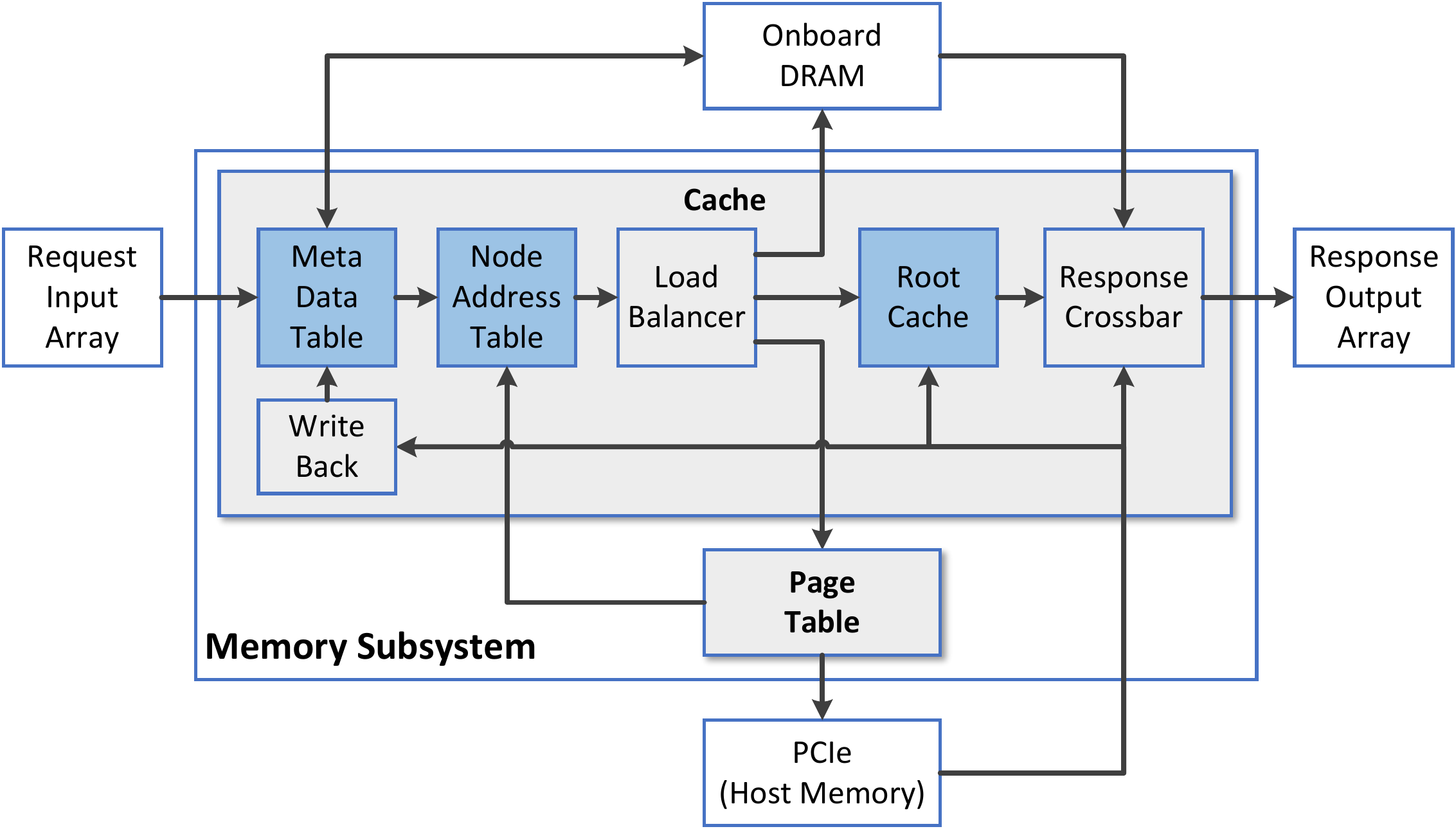}
    \caption{Memory Subsystem}
    \vspace{-8pt}
    \label{fig:mem_subsys_design}
\end{figure}

% the benefits of our cache
As shown in \Cref{fig:mem_subsys_design}, the memory subsystem includes the cache and the page table.
It caches B-Tree nodes to improve performance by reducing off-chip memory accesses, reducing memory accesses over PCIe, and maximizing the use of available off-chip bandwidth. The root node, which is accessed by all requests, is cached in on-chip memory to save off-chip bandwidth. Other interior nodes are cached in on-board DRAM to reduce PCIe accesses, which both improves request throughput and latency. The load balancer directs some memory accesses over PCIe even when they hit in the cache to use all available off-chip memory bandwidth. 

The memory subsystem supports multiple input/output interfaces for parallel access from the interior-node search and leaf-node scan engines. It also
maintains a page table mapping node LIDs to node physical addresses. This is exposed over PCIe for write operations to replace subtrees atomically by changing mappings (\Cref{sec:insert}). When the slow path for a write operation changes a page table mapping, the cache entry for the node with that LID is invalidated. Leaf nodes are not cached to avoid the need for cache invalidations over PCIe for every write operation to the B-Tree.

% metedata table
The metadata table keeps track of cached B-Tree nodes in on-board DRAM. It implements a four-way set associative cache indexed by LIDs. Each entry in a set  records the node LID, the physical address, and a 32-bit occupancy map where each bit represents 256~bytes of node data (with 8~KB nodes). The metadata table itself is stored in on-board DRAM. \sys uses a small on-chip metadata cache, which has capacity for 1K entries in the current implementation, to minimize the traffic to on-board DRAM.

% write back
On cache misses, the cache module fetches an integral number of 256-byte chunks over PCIe and attempts to write them back to the cache. The write back module implements a locking scheme to avoid conflicts when the cache is invalidated on page table updates or during B-Tree root updates. The write back operation allocates space for a new node in the cache only if the missing access was to the header and shortcut block of the node. The write back module fills other chunks in the allocated cache space when it reads sorted block segments for the node.  When interior nodes cannot fit in the cache, we evict a random node from the same set.
We leave more complex cache replacement policies for future investigation. 

% NAT
The metadata table determines if incoming memory read requests are hits or misses, and sends them to the node address table (NAT). The NAT is responsible for ensuring each request has a consistent view of each node it visits. For example, without the NAT, a request could read the shortcut block from the cache and an inconsistent sorted block segment from host memory after the page table entry for the node is updated.
The NAT maps the request sequence number to the physical address of the node version first accessed by the request. The NAT entry can be written by both the metadata table and the page table depending on whether the first access by the request to a node is a hit or a miss. When an access to a sorted block hits in the cache, the physical address in the cache entry must match the  physical address recorded in the NAT for the request. In case of a mismatch or cache miss, the sorted block segment is always loaded from host memory using the physical address from the NAT for the request. 

% load balance

With large caches, the accelerator can become bottlenecked on accesses to on-board DRAM while underutilizing PCIe bandwidth.
To use all available off-chip memory bandwidth, the load balancer directs some memory accesses that hit in the cache to host memory over PCIe. 
It constantly monitors the number of inflight operations and the total number of bytes being read by those operations on both DRAM and PCIe interfaces.
Then balances load across the two interfaces.

\if 0
However, if a high traffic load is diverted to PCIe, its queuing latency can be much higher than that of DRAM. 
The cache hits will firstly consume DRAM throughput, and they will be diverted only when PCIe interface is underutilized. 
\fi

\input{experiments.tex}

\section{Related work}\label{sec:related}

%ordered key value stores
% masstree, ART, bwtree, cuco-trie, openbwtree
We build on a large body of work on indices for in-memory ordered key-value stores (e.g.,~\cite{art,bwtree,masstree,openbwtree,cuckootrie}). A comparison study~\cite{openbwtree} showed that the Adaptive Radix Tree (ART)~\cite{art} performed best followed by Masstree~\cite{masstree} except on scan-dominated workloads where a B+ tree~\cite{btree} variant was better. We chose Masstree~\cite{masstree} as a baseline because it has better scan performance than ART and it supports variable-size keys unlike the B+ tree variant in~\cite{openbwtree}. Cuckoo Trie~\cite{cuckootrie} exploits memory level parallelism to improve multi-core performance but it has worse scan performance than ART. None of these indices supports linearizable scans. Given our focus on scan-dominated workloads, \sys implements a B+ tree variant with support for variable-size keys and MVCC. 
Many of the optimizations in recent work exploit the memory hierarchy of modern server-class CPUs and are not applicable to \sys where the index is accessed over PCIe.

%erpc
%one-sided RDMA based improvements, 
Ordered key-value stores shard an index across servers and provide access to remote clients across the network. Recent research has shown how to leverage kernel-bypass using DPDK~\cite{dpdk} to implement eRPC~\cite{erpc}, a fast RPC mechanism. eRPC and Masstree have been used to implement a high throughput, low-latency, key-value store~\cite{erpc} that we use as our baseline.
Other research has explored using one-sided RDMA reads to bypass the server CPU for {\sc get} and {\sc scan} operations~\cite{farmsosp,drtm,cell,ziegler,learnedcache}. Since RDMA NICs only provide simple reads of contiguous memory, these systems require at least two RDMA reads per operation to support variable-sized keys or values, and they use client-side caching to avoid additional RDMA reads when traversing the index. XStore~\cite{learnedcache} uses a learned cache, 
a compact client-side cache design inspired by the work on learned indices~\cite{learnedindex,ding2020alex}, but it does not support variable-sized keys or linearizable {\sc scan}s.

CliqueMap~\cite{singhvi21cliquemap} implements a hybrid in-memory
key-value caching system with both RPC and remote memory access (RMA). Like \sys, it accelerates {\sc get} with RMA and executes {\sc set} via RPC to save CPU cost and improve performance. However, this implementation is based on a hash table and does not provide support for efficient {\sc scan}s.

There are proposals to extend RDMA, e.g.,~\cite{aguilera2019designing,prism}, but these have yet to be implemented in NIC hardware.

SmartNICs avoid the functionality limitations of RDMA by providing programmable processing engines in the NIC. There are two types: SoC-based with general purpose cores and accelerators for common functionality like compression and encryption (see~\cite{ipipe} for a survey); and FPGA-based~\cite{Putnam2014,Caulfield2016}. Since the later offer the promise of better performance and energy efficiency at the cost of being harder to program, \sys leverages an FPGA-based SmartNIC but implements complex split and merge of B-Tree nodes on the host CPU.

SmartNICs have been used as offloads for accelerating networking, e.g.,~\cite{clicknp,firestone2017vfp,firestone2018azure,floem,hxdp,tonic2020,eran2019nica}, AI inference~\cite{fowers2018a}, distributed file systems~\cite{linefs}, transactions~\cite{schuh2021xenic}, unordered key-value stores~\cite{chalamalasetti2013fpga,kv-direct,ren2019low,eran2019nica}, and ordered key-value stores~\cite{ipipe,yang2021heterokv,heinrich2015hybrid}. 

KV-direct~\cite{kv-direct}, which is the best unordered key-value store offload, stores the index in host memory and
implements hash table reads and writes, which are much simpler than B-Tree writes, in the FPGA.  It achieves better performance than \sys because this reduces synchronization across PCIe and hash table operations only require O(1) memory accesses, but it does not provide {\sc scan}s. The best ordered key-value store offload, HeteroKV~\cite{yang2021heterokv}, implements a B-Tree where the leaves are hash tables, which results in poor scan performance. It cannot support large stores because 
it stores the B-Tree in on-chip FPGA memory, and it supports only fixed-size keys.
\sys provides support for large stores, variable-sized keys and values, fast scans, and strong consistency.

% TODO: say something about iPIPE

%E3 [42] and 𝜆-NIC [25] offload microservices to SoC-based

%btrees on fpga
%yang2021heterokv btree of hashtables. btree on fpga and L2/core sized hash tables as leaves on CPU. Stores the btree nodes in BRAM on-chip memory. So it cannot support very large trees or large keys. A small tree with 4 byte keys and pointers uses 48% of BRAM resources.
% Also only supports fixed-size keys. Do not discuss consistency guarantees and synchronization. When configired to achieve a throughput of 9M scans per %second approximatelly the same as ours, the latency is around 100us. So about 5x worse than ours. Can we say this. If they are returning 100 elements, 
% they are returning more data than us. Rely on batching that works best on small trees. Also they assume that scan do not span leaves. This is not going to %be true in practice, e.g., because the start key is close the boundary at the end of the leaf.

%heinrich2015hybrid and heinrich2018search Only searches the upper levels of the btree in the FPGA and hands the rest of the search to the CPU. Uses the 
% same technique to accelerate the search part of inserts/deletes. They transfer the whole top level of the tree to the FPGA. They only support fixed-size keys
% and they do parallel search across all the keys in a node. Write performance is a disaster and it is assumed that there are almost not updates to the tree. 
% Plus do not look at the cost of transferring the partial results from the FPGA to the CPU over PCIe which dominates latency.

\input{conclusion.tex}

\section*{Acknowledgments}
We thank Vadim Makhervaks, Lukasz Tomczyk, Prahasaran Asokan, and Ankit Agrawal for their help and discussions on exploring Honeycomb in cloud storage services. 
We thank Andrew Putnam for his support and help with onboarding Catapult FPGA platform.
We thank Nikita Lazarev, David Sidler, Mikhail Asiatici and Fabio Maschi who worked on related projects during their internships at Microsoft Research.
We thank the members of Cloud System Futures team in Micrsoft Research Cambridge for their help and feedback.

%%
%% The next two lines define the bibliography style to be used, and
%% the bibliography file.
% \bibliographystyle{ACM-Reference-Format}
\bibliographystyle{plain}
\bibliography{honeycomb_arxiv}

% %%
% %% If your work has an appendix, this is the place to put it.
% \appendix
% \section{XXX}

\end{document}

%% file: abstract.tex
\begin{abstract}
In-memory ordered key-value stores are an important building block in modern distributed applications. We present     
\sys, a  hybrid software-hardware system for accelerating read-dominated workloads on ordered key-value stores that provides linearizability for all operations including scans.
\sys stores a B-Tree in host memory, and executes {\sc scan} and {\sc get} on an FPGA-based SmartNIC, 
and {\sc put}, {\sc update} and {\sc delete} on the CPU. 
This approach enables large stores and simplifies the FPGA implementation but raises the challenge of data access and synchronization across the slow PCIe bus. We describe how \sys overcomes this challenge with careful data structure design, caching, request parallelism with out-of-order request execution, wait-free read operations, and batching synchronization between the CPU and the FPGA. 
For read-heavy YCSB workloads, \sys improves the throughput of 
a state-of-the-art ordered key-value store by at least $1.8\times$.
For scan-heavy workloads inspired by cloud storage, \sys improves throughput by more than $2\times$.
The cost-performance, which is more important for large-scale deployments,
is improved by at least $1.5\times$ on these workloads. 
% For uniform workloads with inserts and more than 80\% {\sc scan} operations, \sys improves the throughput of 
% a state-of-the-art ordered key-value store
% by more than $2.3\times$. It also improves cost-performance, which is more important for large-scale deployments,
% by more than $1.9\times$. 
\end{abstract}

%% file: experiments.tex
\section{Evaluation}\label{sec:eval}
% why not ART-OLC? ART is sensitive to skewness and value distribution of keys. Added text on this to related work
This section compares \sys with eRPC-Masstree \cite{erpc,masstree}, a state-of-the-art ordered key-value store that supports variable-size keys and values (we discuss why we chose this baseline in \Cref{sec:related}). It also evaluates the impact of several optimizations on performance.

% We firstly evaluate the system performance of \sys and eRPC-Masstree on the same server platform. 
% To further understand the differences, we also evaluate the system latency, performance sensitivity on key size and range scan size. 
% Lastly, the impact of our design optimizations are evaluated. 
% These include using log block in leaf nodes, supporting linearizability with MVCC and caching internal nodes in accelerator memory. 

\begin{figure*}[ht!]
    \centering
    \begin{minipage}[b]{0.245\textwidth}
        \centering
        \includegraphics[width=\textwidth]{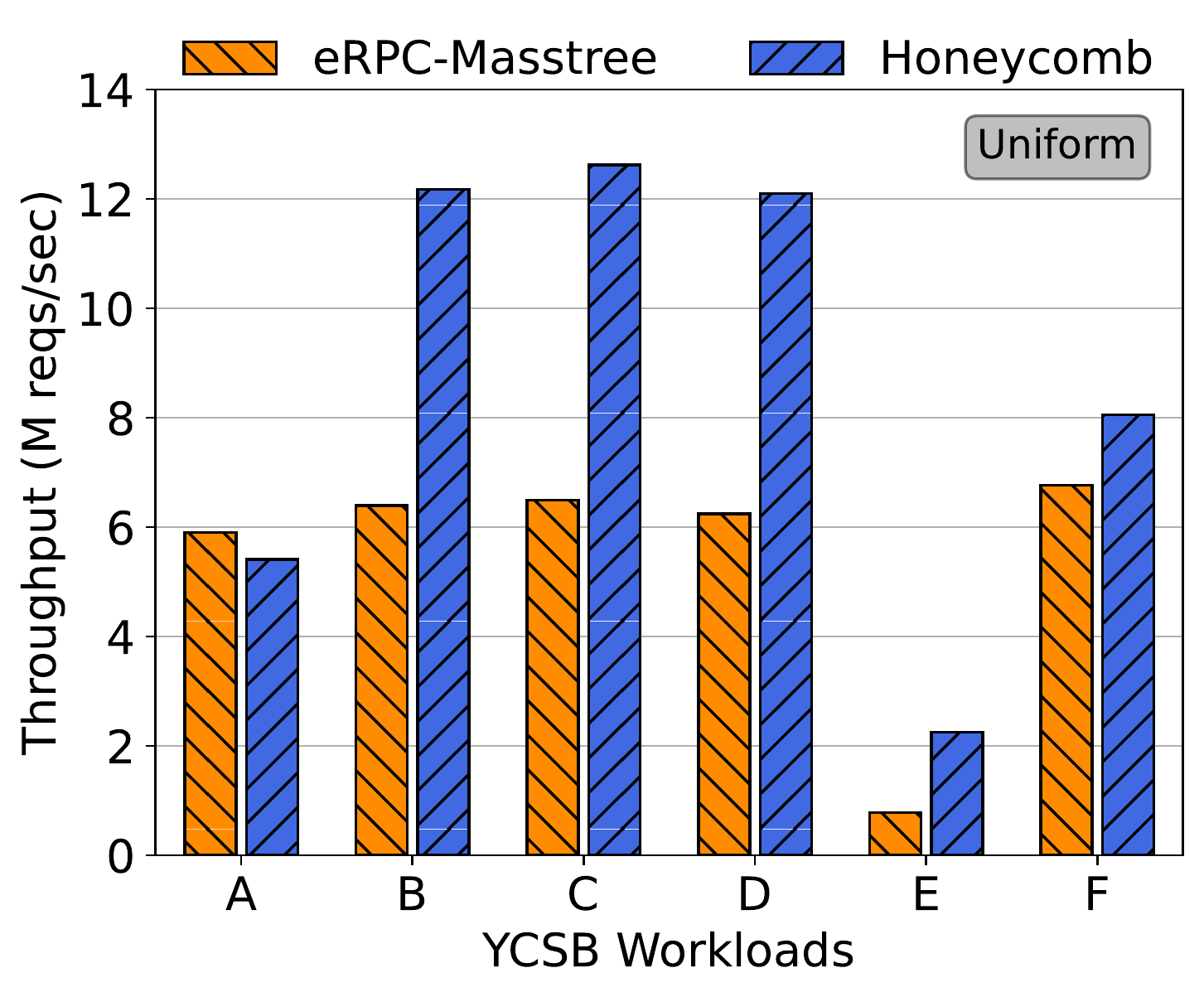}
        % \caption{}
        \label{fig:ycsb_perf_uniform}
    \end{minipage}
    \hfill
    \begin{minipage}[b]{0.245\textwidth}
        \centering
        \includegraphics[width=\textwidth]{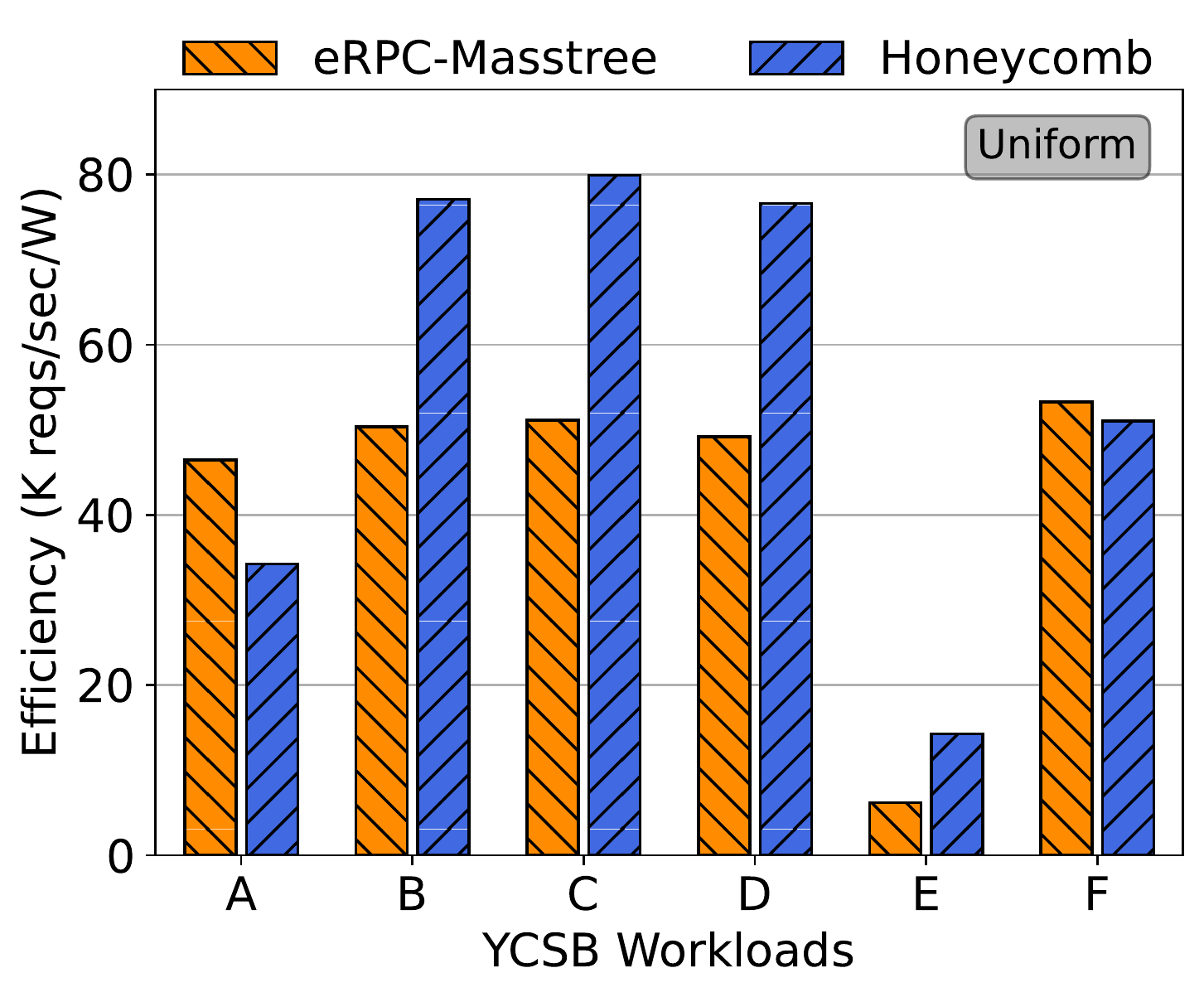}
        % \caption{}
        \label{fig:ycsb_perf_watt_uniform}
    \end{minipage}
    \hfill
    \begin{minipage}[b]{0.245\textwidth}
        \centering
        \includegraphics[width=\textwidth]{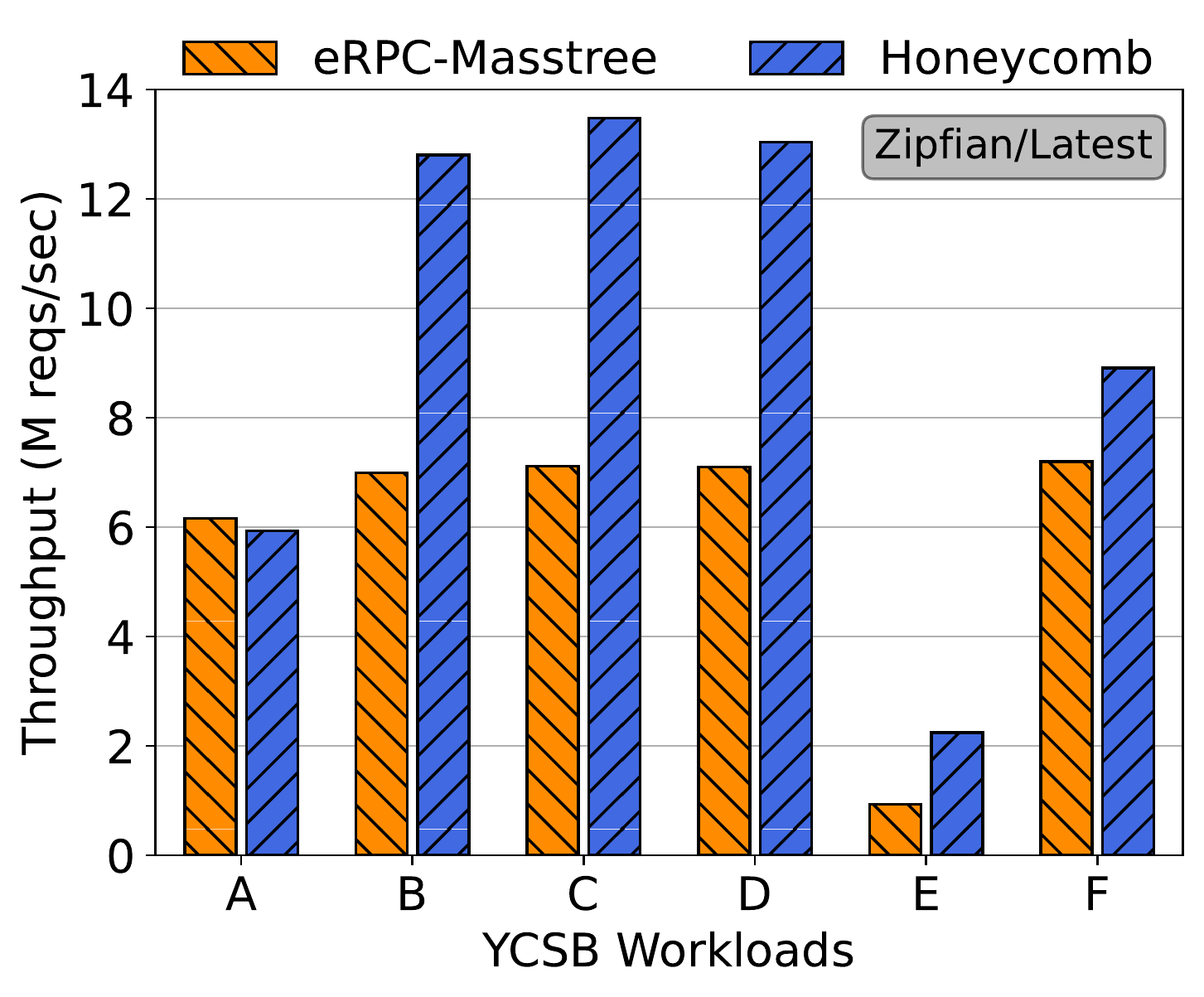}
        % \caption{}
        \label{fig:ycsb_perf_zipf}
    \end{minipage}
    \hfill
    \begin{minipage}[b]{0.245\textwidth}
        \centering
        \includegraphics[width=\textwidth]{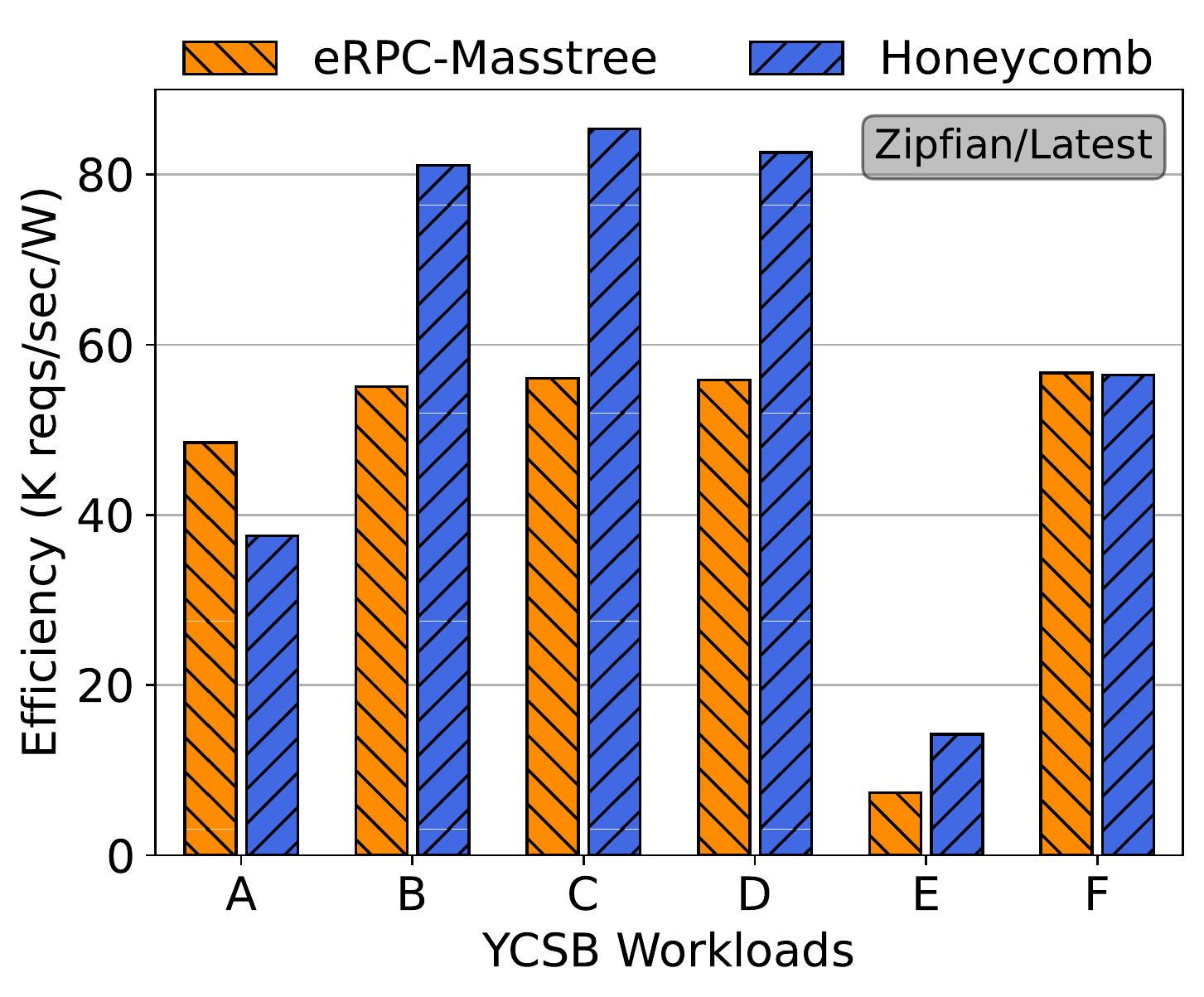}
        % \caption{}
        \label{fig:ycsb_perf_watt_zipf}
    \end{minipage}
    \vskip -16pt 
    \caption{Comparison of throughput and efficiency for YCSB workloads using uniform and Zipfian/latest distributions.}
    \label{fig:ycsb_perf_comparison}
\end{figure*}

\subsection{Experimental setup}
% DELL Z9100-ON switch
% CPU: Intel Xeon E5-2660 v3 @ 2.89 GHz (default 2.60GHZ)
% DRAM: 64GB, 4 channels DDR4-2133
% NIC: Mellanox ConnectX-3
% FPGA: Intel Arria 10 1150 (also used in Microsoft Azure SmartNIC and Brainwave)

Experiments ran one server on one machine and clients on another.
They ran on a single socket of a dual-socket machine because this provided the best cost-performance for both systems. Each
socket had a 10-core Intel Xeon E5-2660 v3 @ 2.89 GHz and four channels to 64~GB of DDR4-2133 DRAM.
The FPGA accelerator card had an Intel Arria 10 1150 FPGA with two channels to 4~GB on-board DDR4-2133 DRAM,
two PCIe Gen3 x8 channels, and 50-Gbps Ethernet. Each machine had 
a separate 40-Gbps Ethernet ConnectX-3 NIC. We used the socket that is attached to both the FPGA and the NIC.
We ran 10 software threads, each pinned to a different core on the socket.
\sys ran on Windows Server 2016 Datacenter using the FPGA, and eRPC-Masstree ran on Ubuntu 20.04 with DPDK 19.11.5 using
the ConnectX-3 NIC.
Machines were connected to a DELL Z9100-ON switch. The different bandwidths did not affect the comparison because 
eRPC-Masstree was never bottlenecked on the network.

\begin{table}[t]
    \centering
    \resizebox{0.9\columnwidth}{!}{
        \begin{tabular}{ l | r r r } 
            \hline
                & Logic & Register & Block RAM \\ 
            \hline\hline
            Shell               & 14.0\%    & 8.9\%     & 14.6\%    \\ 
            Networking          & 6.9\%     & 2.4\%     & 21.0\%    \\
            B-Tree accelerator  & 33.0\%    & 11.5\%    & 35.6\%    \\
            Memory subsystem    & 7.3\%     & 2.9\%     & 7.7\%     \\
            \hline
            Total               & 61.2\%    & 25.7\%    & 78.8\%    \\
            \hline
        \end{tabular}
    }
    \caption{FPGA resource usage of \sys.}
    \label{tab:hw_res_usage}
\end{table}

\begin{table}[t]
    \centering
    \resizebox{\columnwidth}{!}{
        \begin{tabular}{ l | c c c c c c} 
            \hline
                & A & B & C & D & E & F \\ 
            \hline\hline
            \thead{Read \\Operation}      & {\sc lookup}    & {\sc lookup}    & { \sc lookup}    & {\sc lookup}    & \thead{{\sc scan} \\ (1 to 100)}   & {\sc lookup   }  \\
            \thead{Write \\Operation}     & {\sc update}    & {\sc update}    & -                & {\sc insert}    & {\sc insert}   & \thead{{\sc rd-mod-} \\{\sc wr}}  \\
            \thead{Rd-Wr \\Ratio (\%)}    & 50:50     & 95:5      & 100:0     & 95:5      & 95:5     & \thead{50:50 \\ (66.6\% { \sc lookup})}      \\
            \hline
        \end{tabular}
    }
    \caption{YCSB workloads.}
    \label{tab:ycsb_workloads}
\end{table}

We configured \sys with MVCC by default even though eRPC-Masstree does not provide linearizability for scans.
The modular design of the \sys accelerator enables the user to choose configurations with different numbers of 
KSUs, RSUs, and MSIs. This is critical to achieving good power efficiency because it allows the user to trade off performance for power and hardware resource efficiency.
We used a simple analytic performance model and tuning experiments to select the minimum number of these units to achieve a target {\sc scan} operation throughput of approximately 10 Mops/s. We used 14 KSUs, 4 shortcut-RSUs, 5 log-RSUs, 5 sorted-RSUs, and 4 MSI adapters in all the experiments described in this paper. The accelerator is clocked at 220 MHz. The breakdown of FPGA resource usage is shown in \Cref{tab:hw_res_usage}.
The FPGA design tool \cite{quartuspower} reports a TDP of 34.9 W.

We ran each experiment three times and present the average of the results. The range of the results was below 4\% of the average for all experiments.

% removed because it is quite a bit more than half for BRAM. 
%The hardware design only utilizes a bit over half of the entire FPGA. 
%The average logic cost of single KSU/RSU is just $0.7\sim1.2\%$. 

\subsection{Workloads}

We used two sets of workloads in the evaluation {\em YCSB}~\cite{ycsb,ycsbworkloads} and a {\em cloud-storage} workload representative of the distributed file system application~\cite{azurestorage} discussed throughout the paper. We ran all six YCSB workloads (see~\Cref{tab:ycsb_workloads}) with both uniform and skewed access patterns. The read-modify-write operation in YCSB-F is a combination of a {\sc lookup} followed by an {\sc update}. 
The cloud-storage workload is similar to YCSB-E but uses shorter scans and we varied the percentage of scans from 50\% to 100\% to characterize the range of read-write ratios for which \sys is beneficial.  We also varied the number of key-value pairs returned by scans and the size of keys and values. 
In all workloads, insert keys were generated randomly with uniform distribution as in \cite{learnedcache}. We used both uniform and Zipfian~\cite{ycsb} ($\theta=0.99$) distributions for lookup keys and the start keys of scans. 

All experiments used the uniform distribution, and 16-byte keys and values unless specified otherwise. The store had 128 million key-value pairs in the initial state for all experiments, which are stored in a 4-level tree in \sys.
We observed that eRPC-Masstree consumes $3\times$ more memory than the total size of the key-value pairs to store the whole B-Tree, whereas \sys only consumes about $1.44\times$. 

\begin{figure*}[ht!]
    \centering
    \begin{minipage}[b]{0.245\textwidth}
        \centering
        \includegraphics[width=\textwidth]{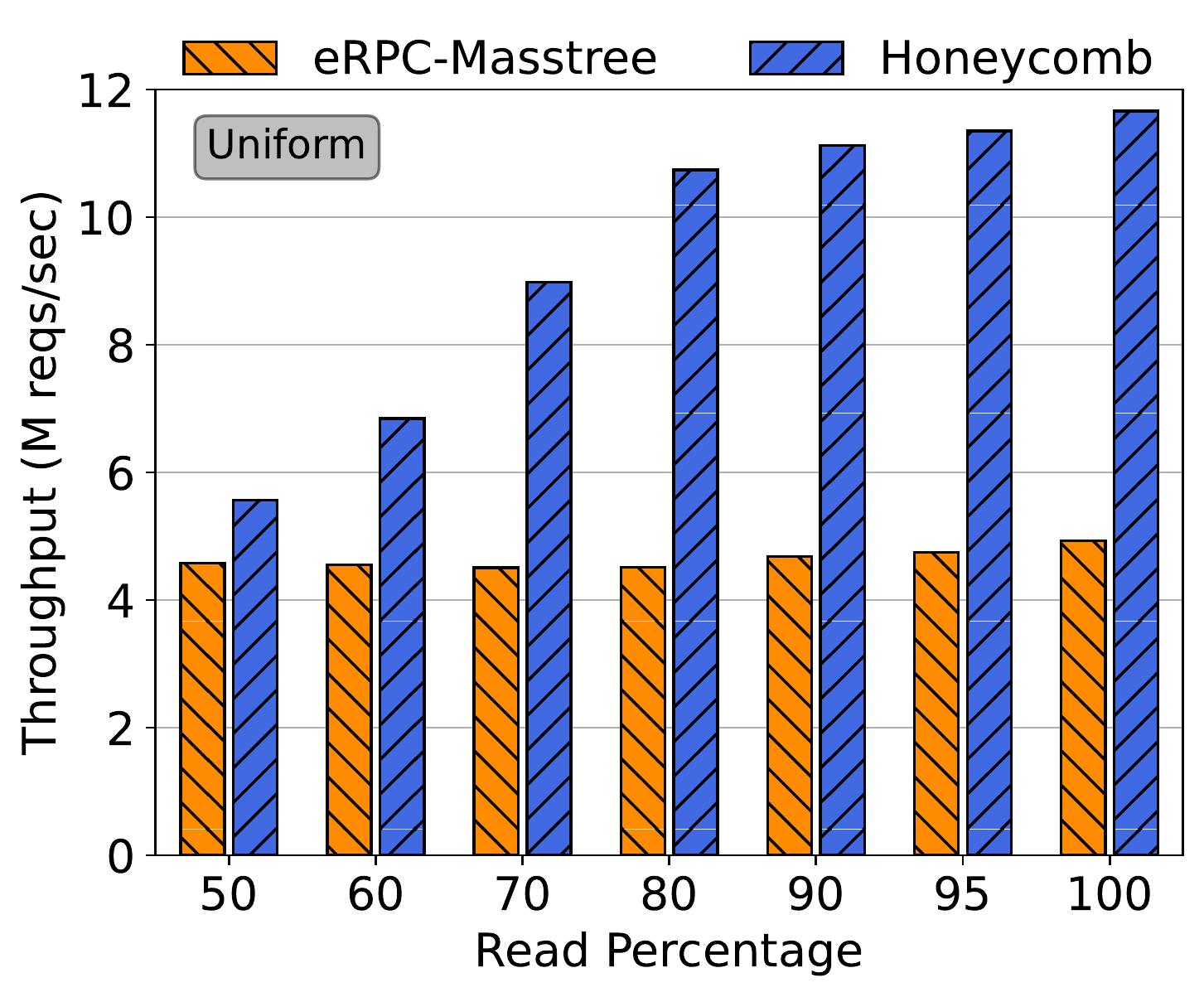}
        % \caption{}
        \label{fig:perf_uniform}
    \end{minipage}
    \hfill
    \begin{minipage}[b]{0.245\textwidth}
        \centering
        \includegraphics[width=\textwidth]{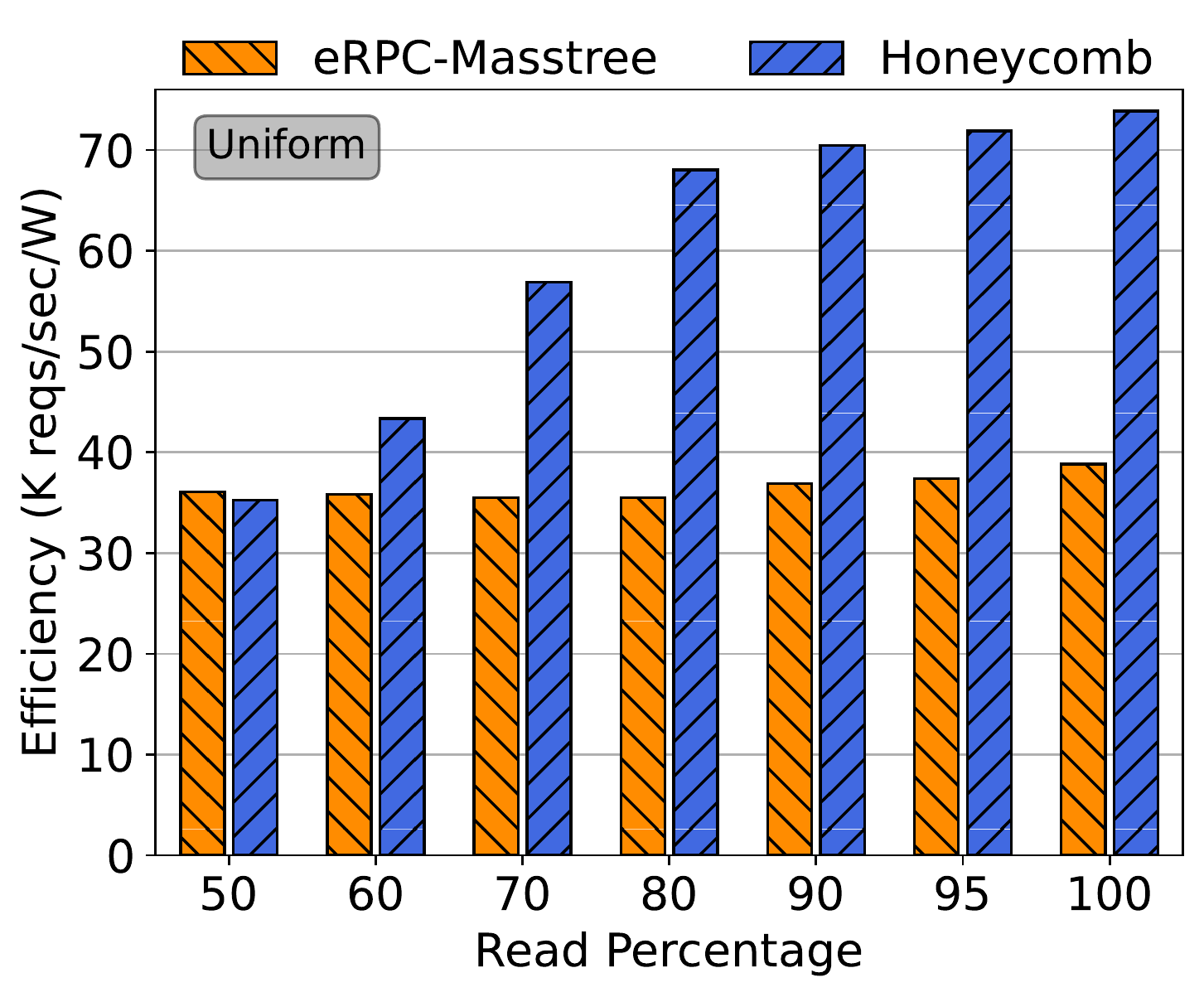}
        % \caption{}
        \label{fig:perf_watt_uniform}
    \end{minipage}
    \hfill
    \begin{minipage}[b]{0.245\textwidth}
        \centering
        \includegraphics[width=\textwidth]{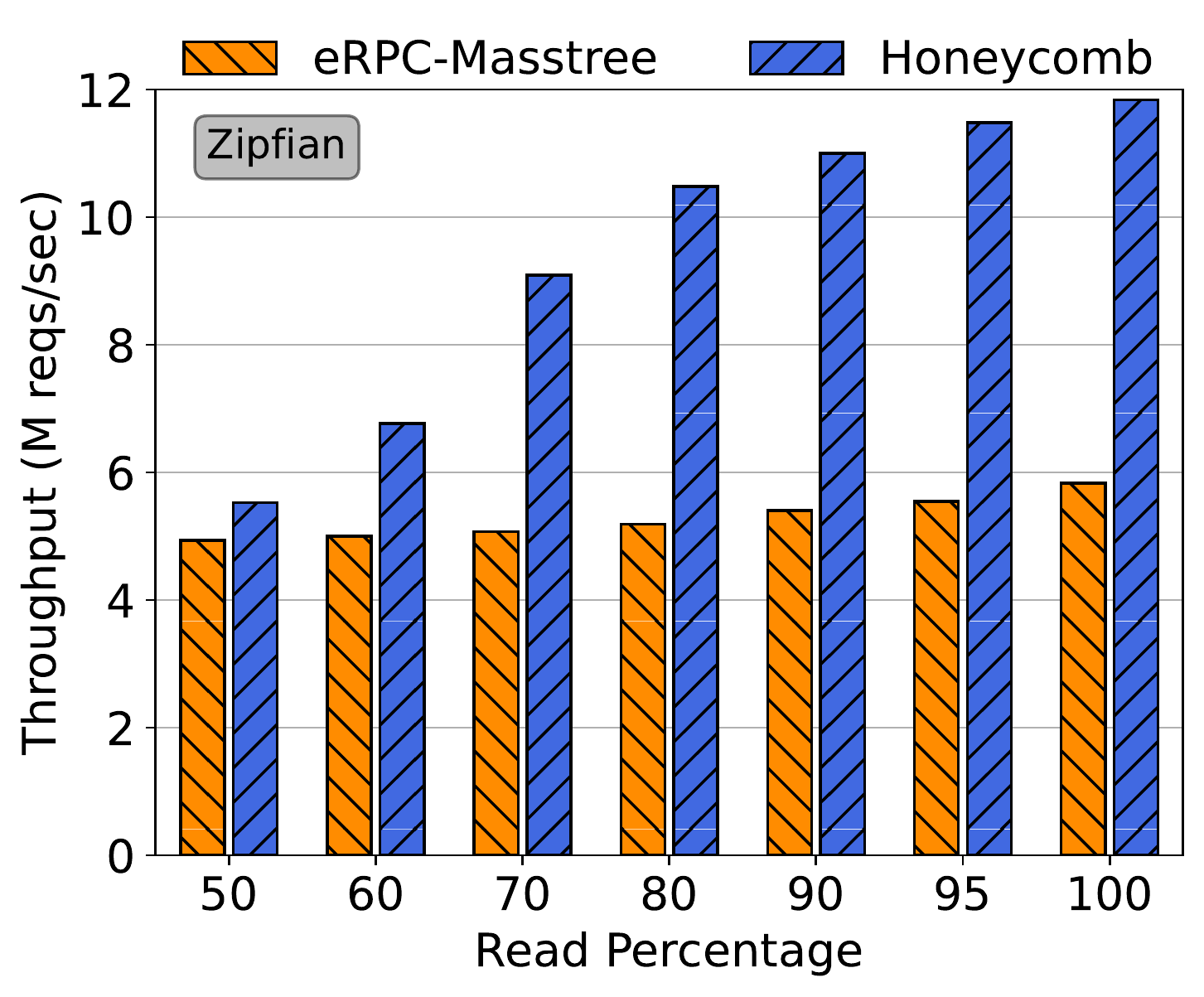}
        % \caption{}
        \label{fig:perf_zipf}
    \end{minipage}
    \hfill
    \begin{minipage}[b]{0.245\textwidth}
        \centering
        \includegraphics[width=\textwidth]{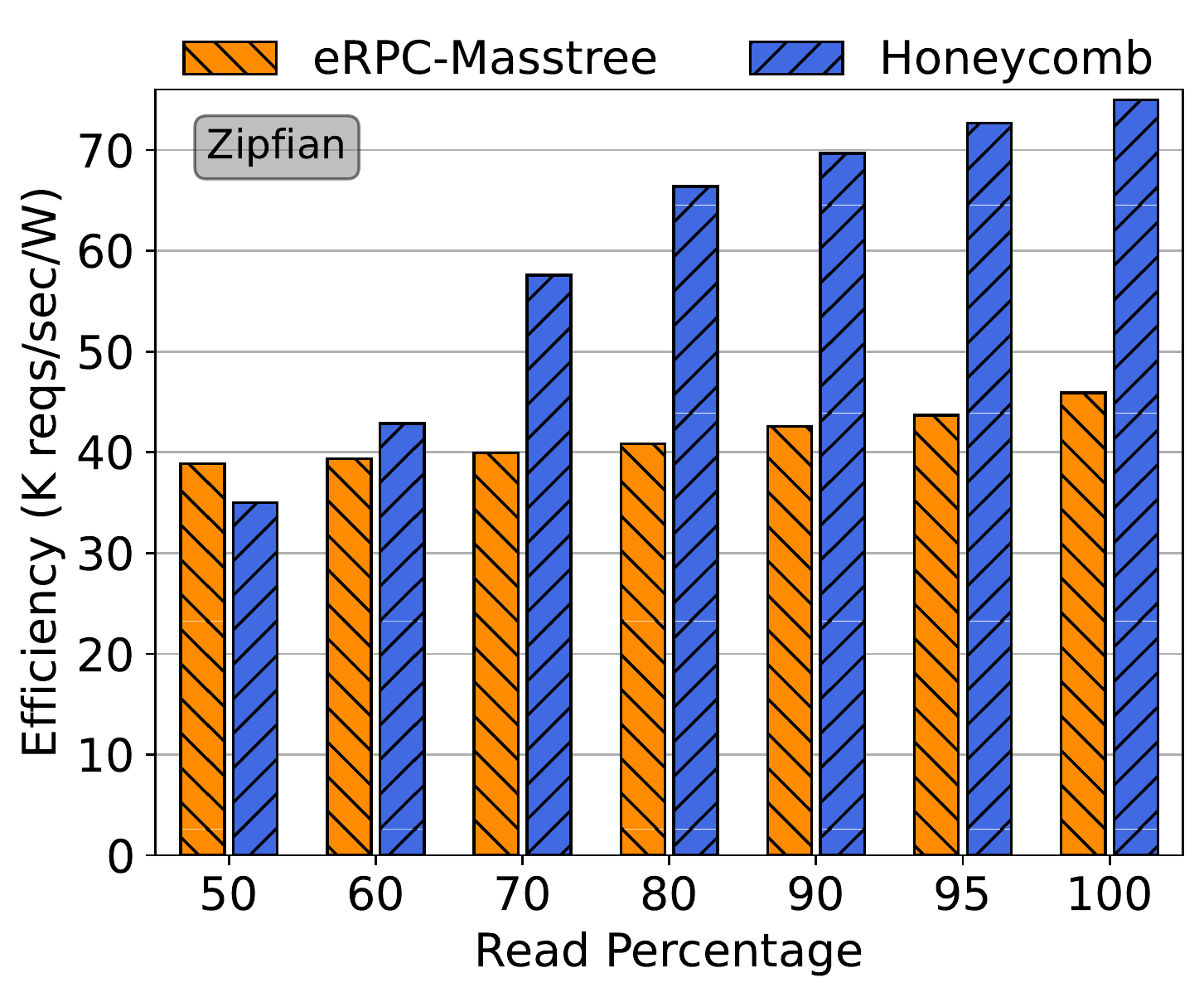}
        % \caption{}
        \label{fig:perf_watt_zipf}
    \end{minipage}
    \vskip -16pt 
    \caption{Comparison of throughput and efficiency for cloud-storage workloads with 50\% to 100\% reads using uniform and Zipfian distributions. The read operation is {\sc scan} with 3 to 4 items in the range. }
    % \vskip -8pt 
    \label{fig:perf_comparison}
\end{figure*}

\begin{figure*}[t]
    \centering
    \begin{minipage}[b]{0.245\textwidth}
    \centering
        \includegraphics[width=\textwidth]{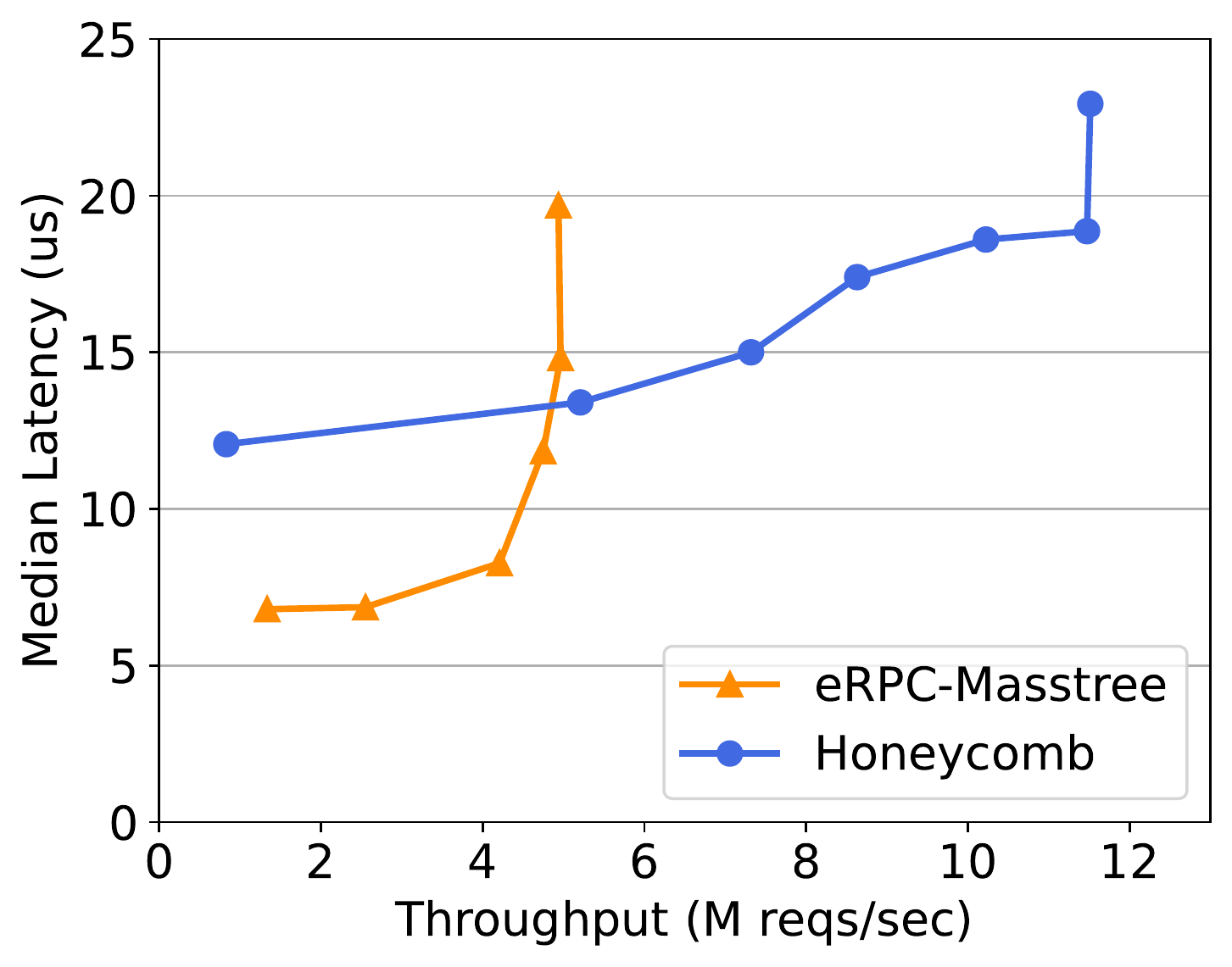}
        \vskip -8pt 
        \caption{Latency-throughput. \\ \quad } % with a uniform distribution
        \label{fig:throughput_latency}
    \end{minipage}
    \hfill
    \begin{minipage}[b]{0.245\textwidth}
      \centering
        \includegraphics[width=\textwidth]{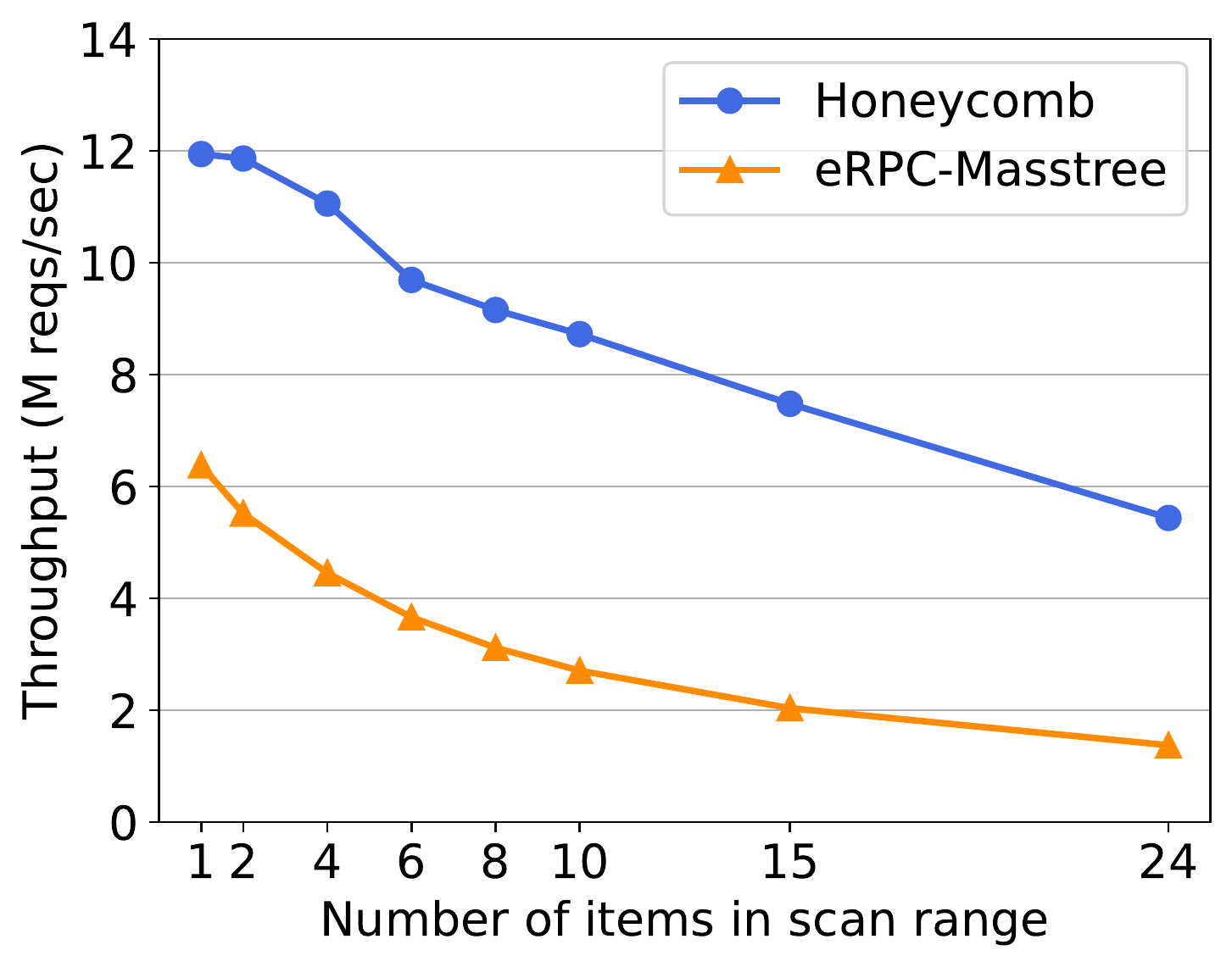}
        \vskip -8pt 
        \caption{Varying {\sc scan} size. \\ \quad }
        \label{fig:scan_size}
    \end{minipage}
    \hfill
    \begin{minipage}[b]{0.245\textwidth}
          \centering
        \includegraphics[width=\textwidth]{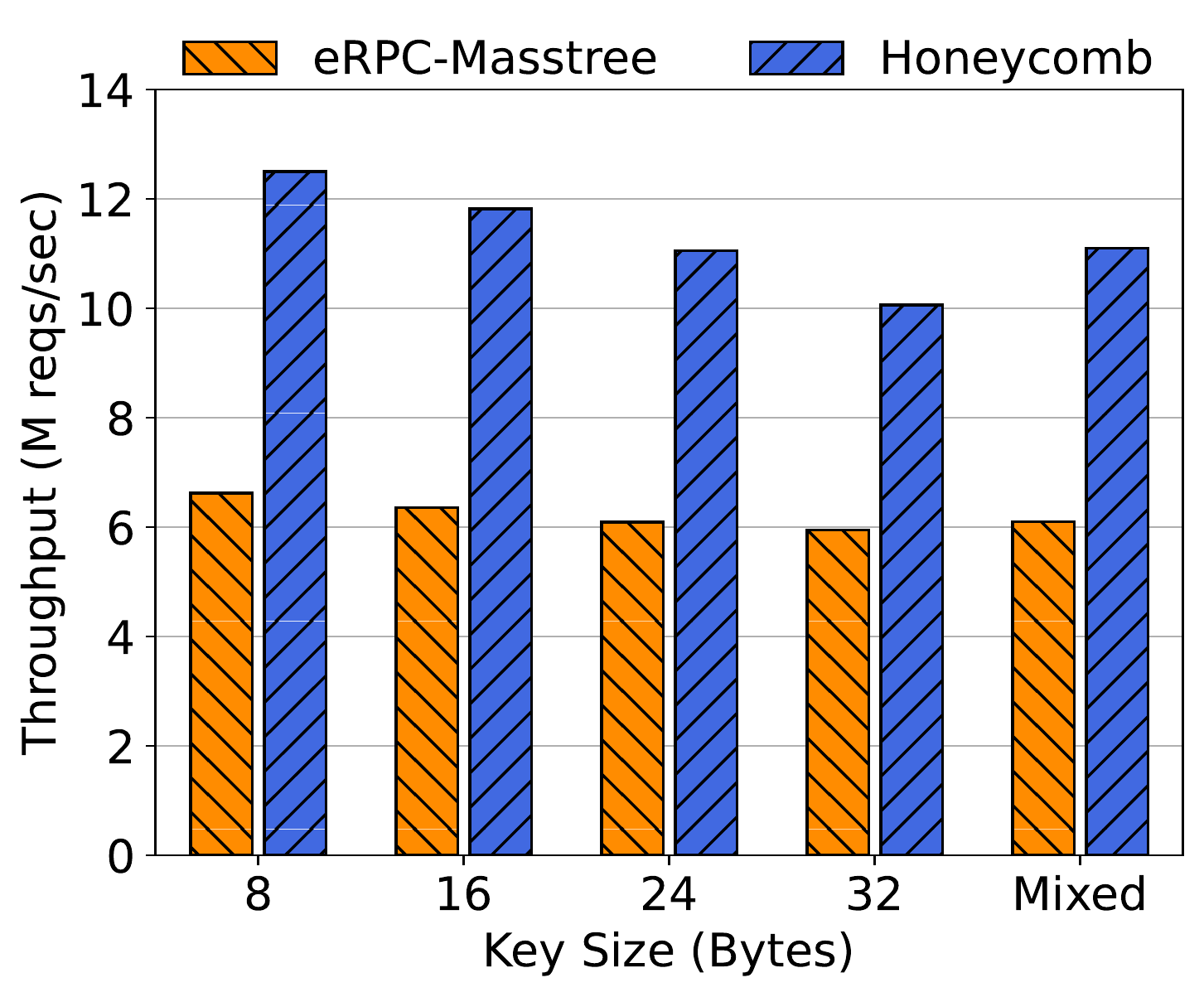}
        \vskip -8pt 
        \caption{Varying key sizes. \\ \quad}
        \label{fig:key_size}
    \end{minipage} 
    \hfill
    \begin{minipage}[b]{0.245\textwidth}
        \centering
        \includegraphics[width=\textwidth]{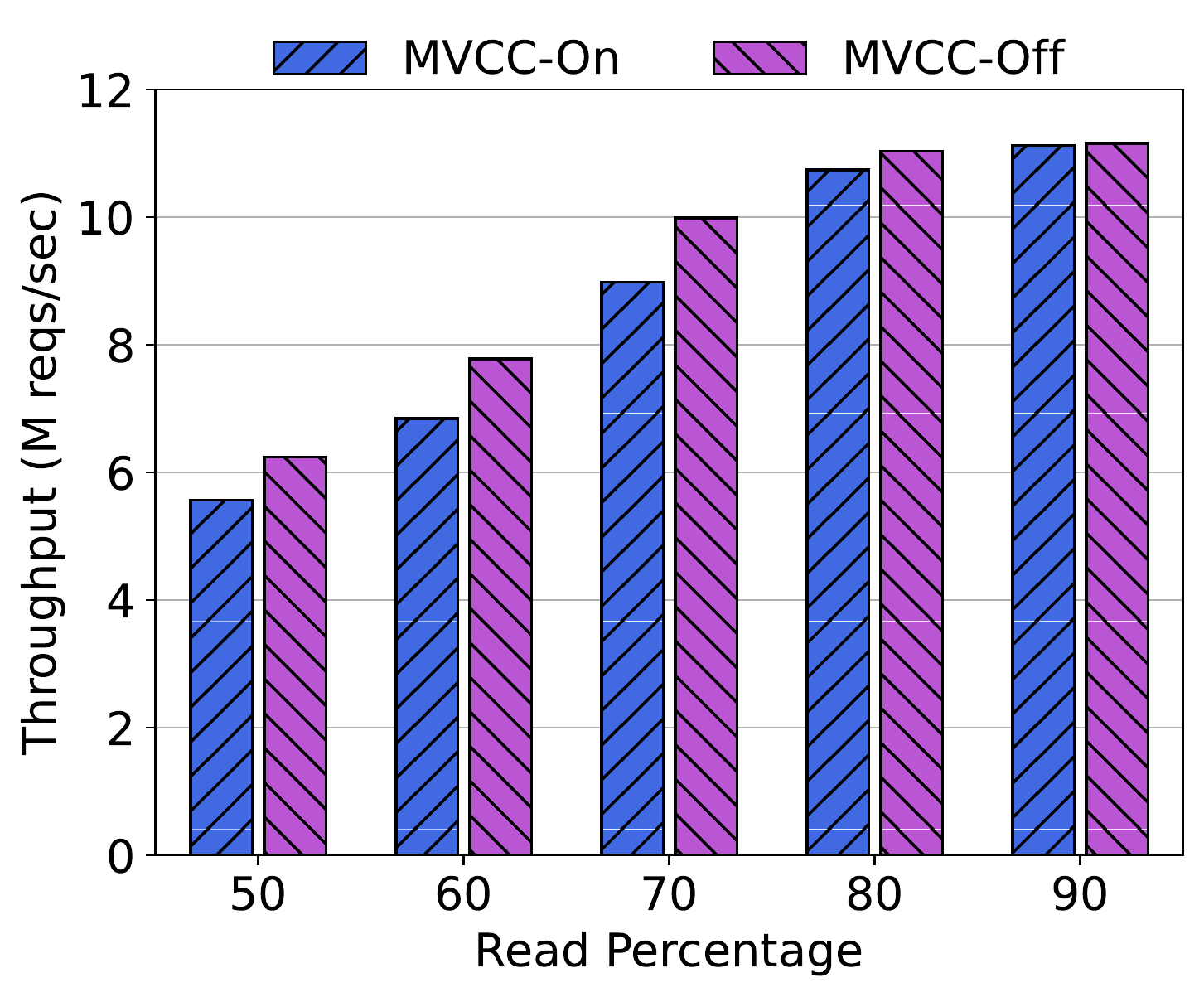}
        
        \vskip -8pt 
      \caption{Impact of MVCC.\\ \quad }
      \label{fig:mvcc_impact}
      \end{minipage}
    \vskip -8pt 
\end{figure*}

\subsection{Comparison with eRPC-Masstree} 

Cost-performance is the key metric to optimize in large scale data centers. We use TDP as a proxy for total cost of ownership (TCO) as in~\cite{jouppi2021ten}. 
%It correlates better with TCO than measured power because it accounts for the cost of provisioning power and cooling when servers run at full power~\cite{jouppi2021ten}. 
We use a single-socket server for \sys because adding another socket increases TDP without increasing the throughput of the hardware accelerator. We ran experiments with two sockets for eRPC-Masstree but they resulted in worse cost-performance, e.g.,  
it achieved $1.34\times$ better cost-performance with one socket than with two for read-only 3-item scan workloads. Therefore, we also present single socket results for eRPC-Masstree. 
We compute TDP by adding numbers for each component from published documentation. The server TDP is 127~W for eRPC-Masstree and 157.9~W for \sys. The TDP for 
\sys is larger because
 it uses a 40-W FPGA accelerator board instead of a 10-W ConnectX-3 NIC \cite{connectx3}.
 
 Both the throughput and the TDP should scale by a factor of two when using an additional CPU socket connected to an additional NIC (for eRPC-Masstree) or SmartNIC (for \sys). Therefore, the performance per watt of TDP for such a system would be similar to the results we present with a single socket.

\if 0
We also ran experiments with a software-only version of \sys running on the CPU. It performed significantly worse than eRPC-Masstree because our design is optimized for hybrid FPGA-CPU execution across PCIe not for exploiting the CPU memory hierarchy. Therefore, we do not show results for the software-only version.
\fi

{\bf YCSB throughput and efficiency:}
\Cref{fig:ycsb_perf_comparison} compares both average throughput (Mreqs/s) and throughput per watt of TDP (Kreqs/s/W) of eRPC-Masstree and \sys.
Since \sys is not optimized for write-heavy workloads, it is less efficient running YCSB-A and F.
% as in YCSB-A~\cite{ycsb}, it achieves similar throughout against eRPC-Masstress by 8\% less with uniform and 4\% less with Zipfian. The corresponding throughput per watt is 26\% and 22\% less respectively. \sys starts to show $1.2\times$  better performance with similar power efficiency as in YCSB-F, where $2/3$ operations are lookups. 
For YCSB-B, C and D, \sys improves throughput per watt by $1.5\times$ with both uniform and Zipfian/latest distributions. It also improves the throughput for these workloads by $1.9\times$ with uniform and $1.8\times$ with Zipfian/latest distributions. 
Since \sys is optimized for scans, it does particularly well in YCSB-E improving throughout by up to $2.9\times$ and efficiency by up $2.3\times$. \sys is bottlenecked on the network for large scans while eRPC-Masstree is always bottlenecked on the CPU.

%As the response message size can be over 3 KB, \sys performance has been throttled when handling large network transfers. Meanwhile, eRPC-Masstress %performance is bottelnecked by CPU.

\begin{figure*}[th!]
    \centering
    % \begin{subfigure}[b]{0.25\textwidth}
    \subfloat[Performance and cache hit rate.]{
        \centering
        \includegraphics[width=0.25\textwidth]{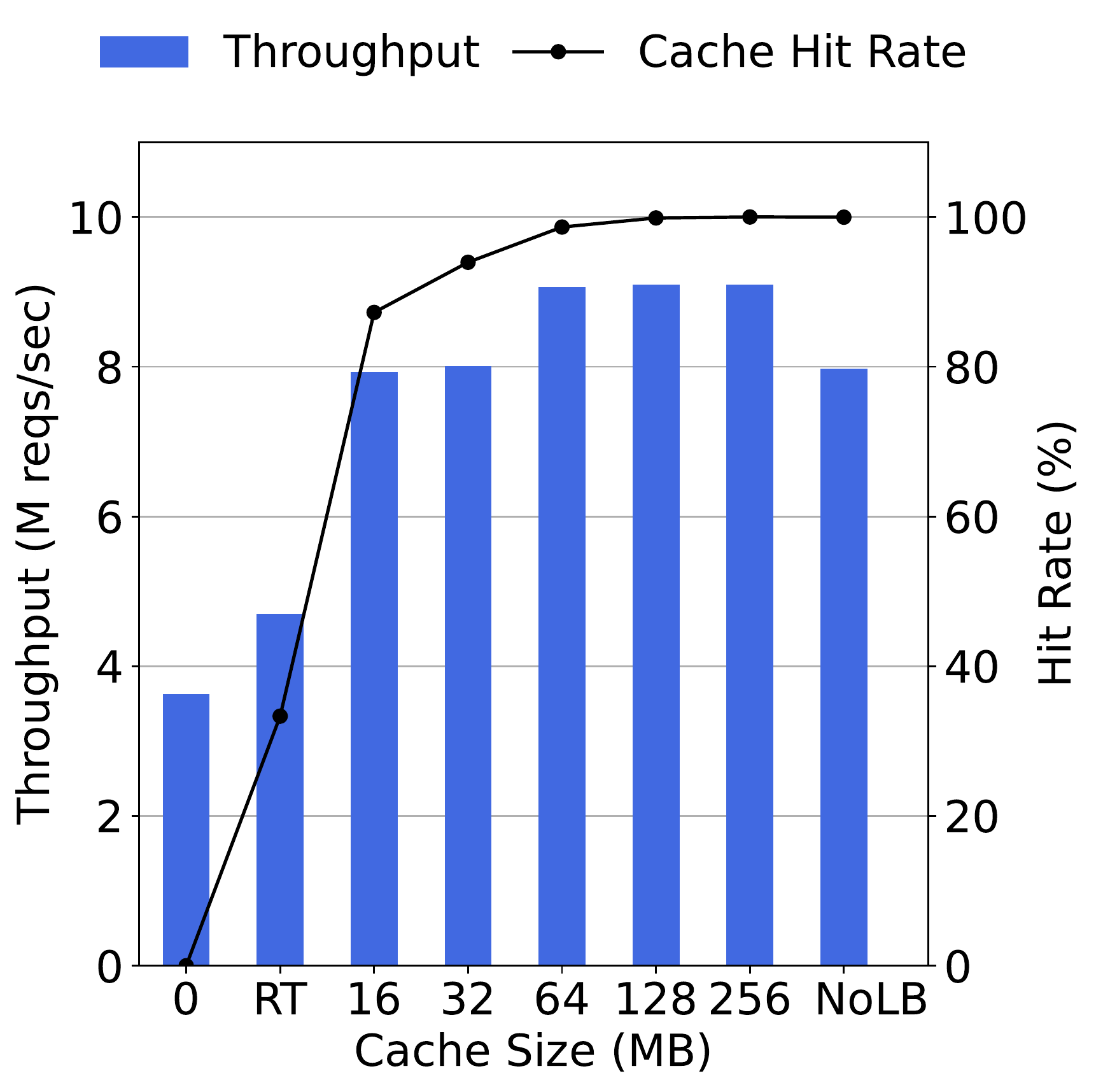}
        % \vskip -8pt 
        % \caption{Performance and cache hit rate. }
        \label{fig:cache_size_tput_hit}
    }
    % \end{subfigure}
    % \hfill
    % \begin{subfigure}[b]{0.37\textwidth}
    \subfloat[FPGA memory bandwidth breakdown.]{
        \centering
        \includegraphics[width=0.36\textwidth]{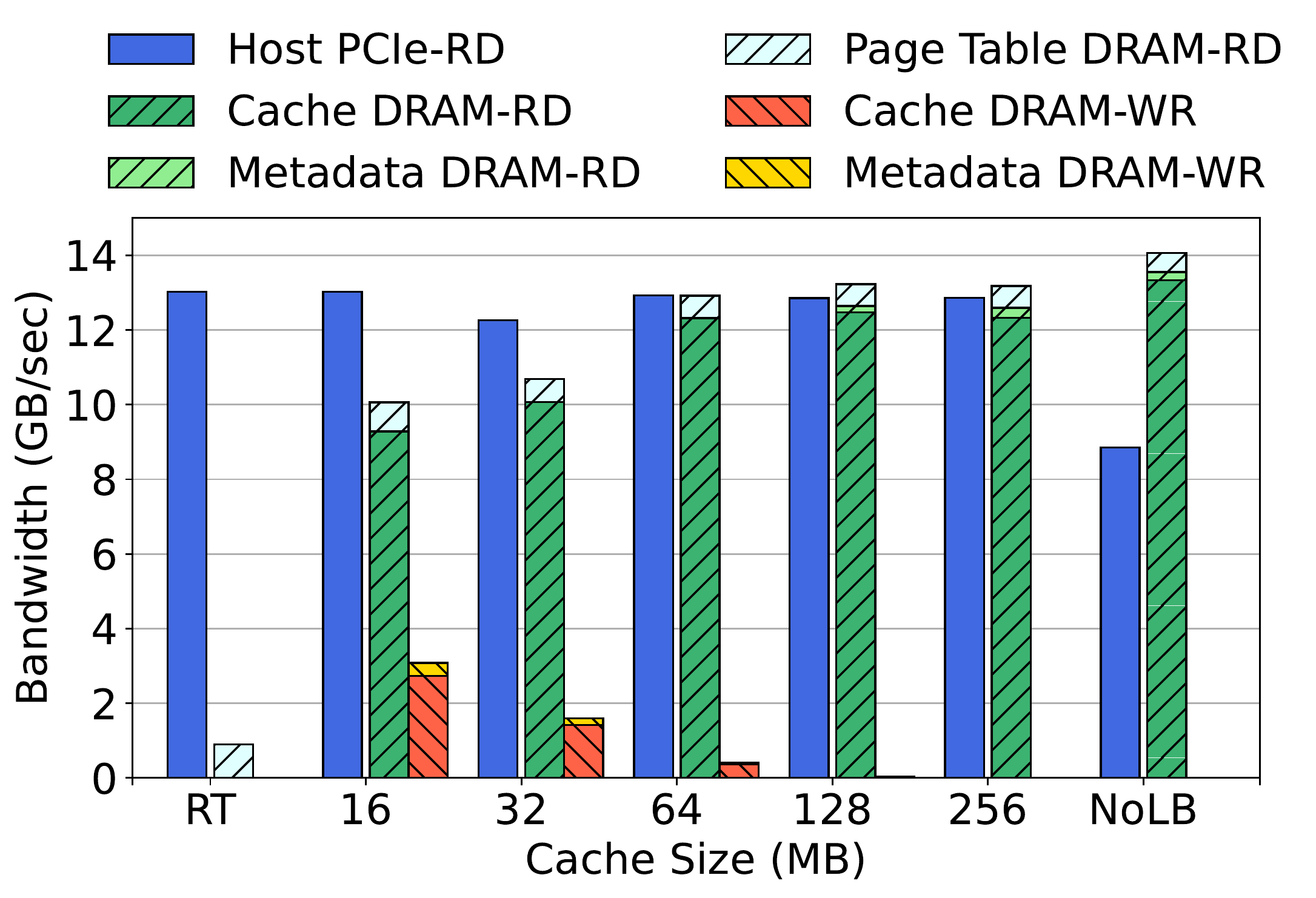}
        % \vskip -8pt 
        % \caption{FPGA memory bandwidth breakdown. }
        \label{fig:cache_size_bandwidth}
    }
    % \end{subfigure}
    % \hfill
    % \begin{subfigure}[b]{0.37\textwidth}
    \subfloat[FPGA memory IOPS breakdown.]{
        \centering
        \includegraphics[width=0.36\textwidth]{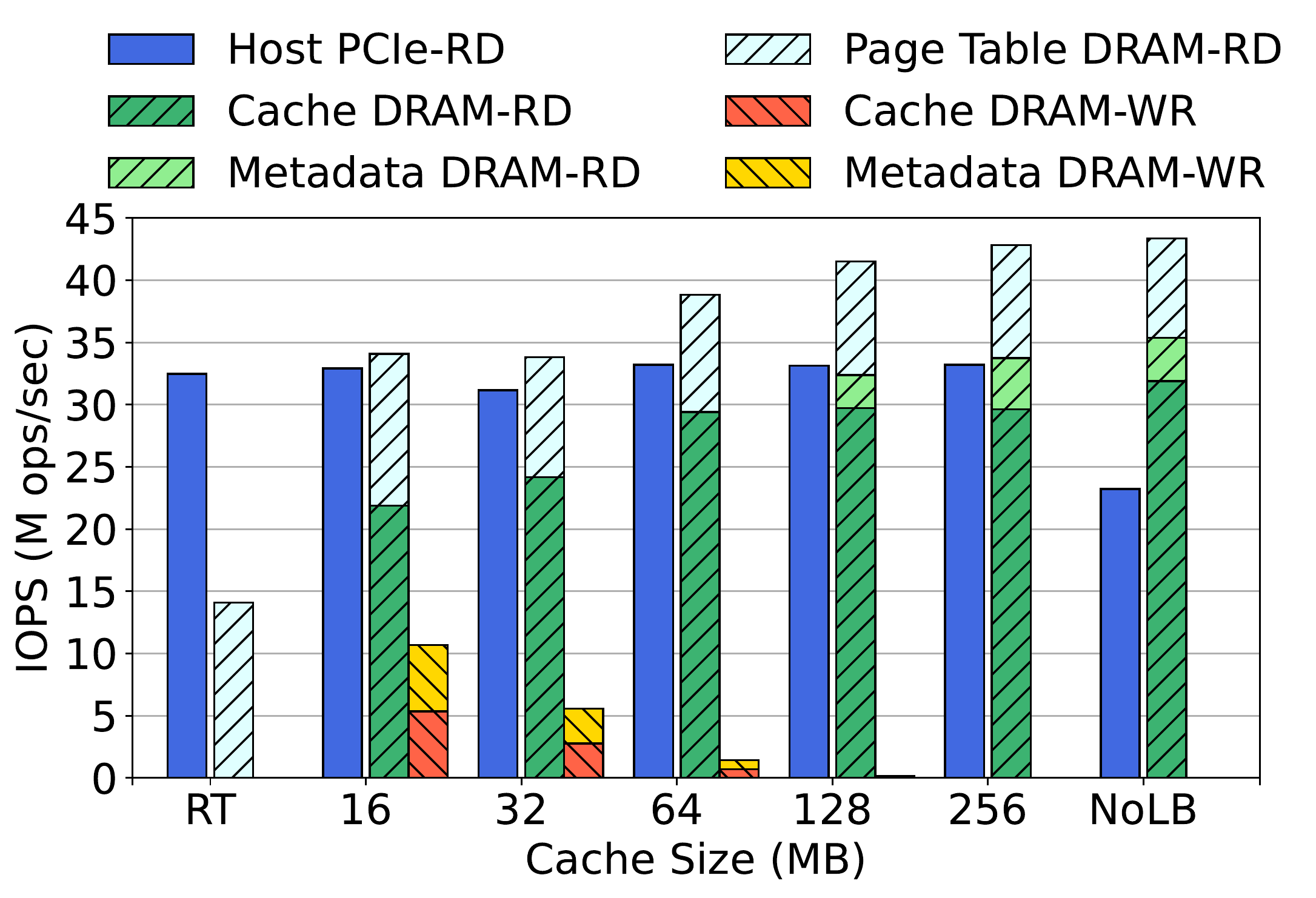}
        % \vskip -8pt 
        % \caption{FPGA memory IOPS breakdown.}
        \label{fig:cache_size_iops}
    }
    % \end{subfigure}
    %%
    \vskip -6pt 
    \caption{Performance impact of caching and load balancing on \sys (FPGA only) 1-item {\sc scan}. }
    \vskip -6pt 
    \label{fig:cache_size}
\end{figure*}

\begin{figure}
    \centering
          \includegraphics[width=0.7\columnwidth]{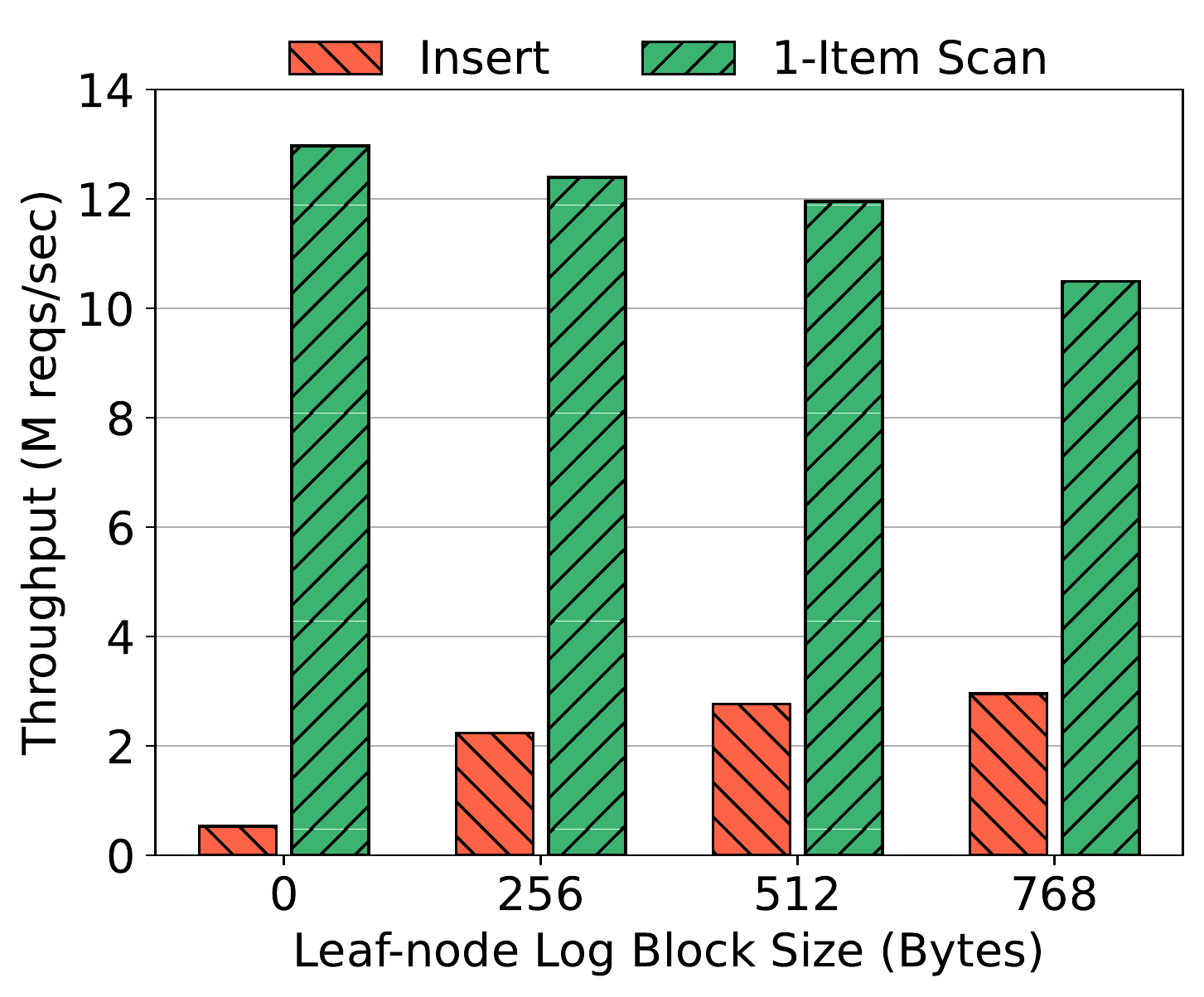}
          \vskip -8pt 
          \caption{Performance impact of log block size. }
          \vskip -6pt 
          \label{fig:log_block_size}
\end{figure}

{\bf Cloud storage throughput and efficiency:} \label{sec:cloud_storage_compare}
% \Cref{fig:perf_comparison} compares both average throughput (Mreqs/s) and throughput per watt of TDP (Kreqs/s/W) of eRPC-Masstree and \sys. 
\Cref{fig:perf_comparison} compares performance and efficiency of eRPC-Masstree and \sys running the cloud-storage workload.
In this experiment, the boundary keys in \sys~{\sc scan(K$_l$, K$_u$)} requests were chosen to return exactly three items when executed on the inital key-value store. However, they can return between three and four items because of newly-inserted items. Since eRPC-Masstree provides a different {\sc scan(K,N)} operation that returns the {\sc N} items following {\sc K}, we selected a value of $N$ between three and four to ensure {\sc scan} operations return the same average number of items in both systems. For high read percentages, \sys also uses CPU cores to execute {\sc scan} operations. We configured eRPC-Masstree to enable any thread to execute {\sc scan}s.

The results show that \sys improves both performance and cost-performance significantly for scan-mostly workloads. Since \sys accelerates only read operations, the improvement grows with the read percentage. In the worst case of 100\% writes, \sys achieves only 58\% of the throughput of eRPC-Masstree (47\% of the cost-performance) because there is no hardware acceleration and the B-Tree is optimized for hybrid CPU-FPGA execution. However, even for workloads with 50\% writes, \sys achieves a similar throughput per watt as eRPC-Masstree and better throughput ($1.2\times$ better with uniform distribution). For workloads with at least 80\% reads, \sys improves throughput per watt by $1.9\times$ with uniform and $1.6\times$ with Zipfian distributions. It also improves the throughput for these workloads by $2.3\times$ with uniform and $2.0\times$ with Zipfian distributions.

% We also ran YCSB D (95\% gets and 5\% inserts) on both systems. \sys improves throughput per watt  by 1.5$\times$  with uniform and 1.4$\times$ with Zipfian. It also has 1.9$\times$ better throughput with uniform and 1.7$\times$ better with Zipfian.

As in YCSB workloads, the gains from acceleration are lower with the Zipfian distribution. Whereas eRPC-Masstree can leverage better locality with CPU caching, the current implementation of \sys does not cache leaf nodes, which prevents it from caching leaves containing popular items. We plan to explore leaf caching in the future.

With modern cloud storage server designs that leverage NVMe SSDs~\cite{ocp2021nvme, ocpSamsungSSD} and fast networks to provide tens of millions of IOPS per server, indexing metadata requires powerful CPUs that account for a large fraction of the overall TCO, e.g., CPUs, DRAM, and the NIC account for half the TDP in Open Compute Project's Poseidon~\cite{ocp2022poseidon} storage server. Therefore, \sys can significantly increase overall performance per TCO for these storage servers, e.g., we estimate improvements around 20\% for server designs similar to Poseidon. This is very significant for large-scale data center deployments.

We expect \sys's cost-performance gains to increase with future hardware because newer FPGAs have more on-chip memory to cache B-Tree nodes and use PCIe Gen5 that has $4\times$ the bandwidth of PCIe Gen3.  Despite the bandwidth improvement, we expect PCIe to remain the bottleneck because we can use configurations with more KSUs, RSUs, and MSIs to increase parallelism. Therefore, the techniques proposed in this paper will continue to be important.

{\bf End-to-end latency:}
\Cref{fig:throughput_latency} shows throughput-latency curves for median latency measured at the client in a read-only workload where each {\sc scan} returns exactly three items. \sys can provide better throughput than eRPC-Masstree's throughput at lower latency. However, eRPC-Masstree has lower latency at low load. This is mostly because \sys memory accesses over PCIe and to on-board DRAM have higher latency than CPU memory accesses.
%  but also because the network path in the ConnectX-3 NIC is faster than the one implemented in the FPGA.
%TODO: explain why? some of the network path in soft logic and also PCI dma

{\bf Scan size:}
\Cref{fig:scan_size} shows the throughput of both systems for a read-only workload when varying the number of items returned by {\sc scan}s.
The gains of acceleration increase with scan size, for example, \sys has $4.0\times$ better throughput and $3.2\times$ better throughput per watt with 24-item scans.
Since eRPC-Masstree must follow pointers to each item in the scan range, these random memory accesses become a bottleneck.
\sys can amortize PCIe accesses to a leaf node over many items because it inlines variable-sized items in leaves. 
% We limit {\sc scan} length to 24 because the current implementations of both \sys and eRPC-Masstree limit responses to 1KB.
With longer scans \sys becomes network bound while eRPC-Masstree remains CPU bound (as observed in YCSB-E).

%We evaluate {\sc scan} length up to 24, with  (nearly 1-KB response message ) because \sys consumes close to 40-Gbps network bandwidth. 
%When scanning further more items, \sys performance will be network bound.  

% Since \sys implements {\sc get} using {\sc scan}, the result for 1-item {\sc scan} corresponds to YCSB C.
% We ran YCSB C on eRPC-Masstree. It achieves 2.6\% higher throughput with {\sc get} than with {\sc scan}. \sys
% achieves 1.8$\times$ better throughput and 1.5$\times$ better throughput per watt on YCSB C.

{\bf Key size:}
\Cref{fig:key_size} shows the throughput of both systems running 1-item {\sc scan} on the initial store when the size of keys and values increases (equal key and value sizes). We use 1-item {\sc scan} to better isolate the impact of increasing key sizes on tree traversal.
The performance of both systems drops as key size increases. eRPC-Masstree has deeper trees to traverse with larger keys.
The depth of the \sys does not change but the accelerator must fetch larger sorted block segments.
We also compared both systems on a store with a mix of key and value sizes chosen uniformly from multiples of 8B less than or equal to 32B.
These demonstrate that \sys performs well with variable-sized KV pairs. 
% Their performance is similar to accessing the store with 24B-sized KV pairs. 
% and the time to compare keys in KSUs and RSUs also increases.
%Ideally, we can further increase the number of processing units to avoid this bottleneck. 

\if 0
{\bf Tree size:}
% \Cref{fig:tree_size} shows 
The throughput of both systems running 1-item {\sc scan} on a store with 16-byte keys and values when the number of key-value pairs increases. 
The performance of eRPC-Masstree drops by up to 7.8\%, while \sys sustains the same throughput with the same B-Tree depth. 
We also observe that eRPC-Masstree consumes $3\times$ more memory than the total size of the key-value pairs, whereas \sys only consumes about $1.44\times$. 
\fi

\subsection{Impact of optimizations}

We ran experiments to investigate the performance impact of using MVCC, log blocks, the cache, and the load balancer.

{\bf Cost of MVCC:}
All experiments ran with MVCC to provide linearizable scans. Since eRPC-Masstree does not provide linearizable scans, we also ran experiments to evaluate the impact of MVCC on throughput.
\Cref{fig:mvcc_impact} shows that turning off MVCC improves \sys performance on cloud-storage workloads by up to 14\% when the workload is bottlenecked by {\sc insert}s. The overhead on read-heavy workloads is negligible. 

%As described in \Cref{sec:rel_ring}, the release ring improves write performance by batching multiple updates to the global read version in the accelerator instead of updating it on every write. 
%The results show that 
% The results show that the release ring significantly improves throughput. It reduces MVCC overhead to 7\% for workloads with at least 90\% reads. We intend to improve performance further by allowing threads to run other operations instead of blocking to release changes, while delaying responses to clients until changes are released to preserve consistency.

{\bf Log block:}
\Cref{fig:log_block_size} shows throughput of 1-item scans in a read-only workload and of inserts in a write-only workload with varying log block size.
Using log blocks improves insert performance because it avoids adding the new item to the sorted block, adjusting shortcuts, and updating the accelerator page table on every operation. However, it reduces {\sc scan} performance because it increases the amount of data accessed over PCIe. Using a 512-byte block achieves most of the benefit while decreasing {\sc scan} throughput by only 8\%. 

{\bf Accelerator cache and load balancer:}
\Cref{fig:cache_size} shows the impact of caching on 1-item {\sc scan} throughput  in a read-only workload using only the FPGA (no CPU).
\Cref{fig:cache_size_tput_hit} shows request throughput and cache hit rate for interior node accesses.
Caching the root node in on-chip memory (RT) improves performance by 30\% over the no-cache case.
Adding an on-board DRAM cache with 256~MB (which can hold all interior nodes) improves performance by 
$2.5\times$. The load balancer directs some cache hits to host memory over PCIe
to maximize off-chip bandwidth utilization. Removing the load balancer (NoLB) decreases throughput by 13\%.

Figures \ref{fig:cache_size_bandwidth} and \ref{fig:cache_size_iops} break down bandwidth and IOPS for PCIe and on-board DRAM traffic.
For small cache sizes, the bottleneck is PCIe bandwidth and there are writes to on-board DRAM due to cache replacement. With a 256~MB cache
(100\% hit rate) and no load balancing (NoLB), on-board DRAM bandwidth is the bottleneck while 4 GB/s of PCIe bandwidth is left unused.
The load balancer shifts traffic to PCIe to maximize off-chip bandwidth utilization increasing throughput from 8.0 to 9.1 Mreqs/s.
Page table and cache metadata reads consume a small fraction of available on-board DRAM bandwidth but a significant fraction of IOPS. For small cache sizes, there are no metadata reads from on-board DRAM because all metadata fits in the on-chip cache.

%% file: conclusion.tex
\section{Conclusion}

In-memory ordered key-value stores are an important building block in modern distributed applications. We presented \sys, a system that leverages an FPGA-based SmartNIC to accelerate these stores with a focus on scan-dominated workloads. It stores a 
B-Tree in host memory, and executes {\sc scan} and {\sc get} operations on the FPGA, and {\sc put}, {\sc update}, and {\sc delete} operations on the CPU.
This approach enables large stores and simplifies the FPGA implementation but raises the challenge of data access and synchronization across slow PCIe. We described how \sys addresses this challenge by using large B-Tree nodes with shortcuts; caching in on-chip and on-board FPGA memory;  exploiting request-level parallelism with out-of-order execution; making {\sc scan} and {\sc get} operations wait free; and using a log in B-Tree nodes to batch synchronization across PCIe.
\sys is evalutated against a state-of-the-art ordered key-value store using both YCSB and a cloud-storage-inspired workload.
The comparison shows that \sys 
improves throughput by $1.9\times$ for uniform read-dominated workloads in YCSB and by $2.3\times$ for uniform cloud-storage-inspired workloads with more than 80\% {\sc scan} operations. Most importantly, \sys improves cost-performance, which is the key metric to optimize in large-scale data centers;  it improves throughput per watt of TDP (a proxy for TCO) by more than $1.5\times$ and $1.9\times$ for these two set of workloads respectively.